\documentclass[%
preprint,
amsmath,amssymb,
aps,
]{revtex4-1}

\usepackage{graphicx}
\usepackage{dcolumn}
\usepackage{bm}

\usepackage{color}
\begin{document}


\title{Thermodynamics of the Quark-Gluon Plasma within a $\bm T$-matrix approach}

\author{Gwendolyn \surname{Lacroix}} 
\email[E-mail: ]{gwendolyn.lacroix@umons.ac.be}
\author{Claude \surname{Semay}}
\email[E-mail: ]{claude.semay@umons.ac.be}
\affiliation{Service de Physique Nucl\'{e}aire et Subnucl\'eaire, Institut Complexys,
Universit\'{e} de Mons -- UMONS, Place du Parc 20, 7000 Mons, Belgium}

\author{Fabien \surname{Buisseret}}
\email[E-mail: ]{fabien.buisseret@umons.ac.be}
\affiliation{Service de Physique Nucl\'{e}aire et Subnucl\'{e}aire, Institut Complexys,
Universit\'{e} de Mons -- UMONS,
Place du Parc 20, 7000 Mons, Belgium;\\ 
Haute \' Ecole Louvain en Hainaut (HELHa), Chauss\'ee de Binche 159, 7000 Mons, Belgium}

\date{\today}

\begin{abstract}
The strongly-coupled phase of the quark-gluon plasma (QGP) is studied here by resorting to a $T$-matrix formulation in which the medium is seen as a non-ideal gas of quasiparticles (quarks, antiquarks and gluons) interacting nonpertubatively. In the temperature range under study, (1-5) $T_c$, where $T_c$ is the temperature of deconfinement, the interactions are expected to be strong enough to generate bound states. The dissociation temperature of such binary bound states is thus computed here. The more the quasiparticles involved in the binary system are heavy, the more the bound state is likely to survive significantly above $T_c$. Then, the QGP equations of state at zero and small baryonic potential are computed for $N_f = 2$ and $N_f = 2 + 1$ by resorting to the Dashen, Ma and Bernstein formulation of statistical mechanics. Comparisons with current lattice QCD data are presented. 

\end{abstract}


\maketitle


\section{Introduction}

\par The phenomenology related to the QCD confinement/deconfinement phase transition is nowadays a fascinating subject in the center of intense investigations, both experimentally and theoretically (see \textit{e.g.}~\cite{yagi} for a review of the topic). During the last two decades, a new state of matter, the quark-gluon plasma (QGP), has been intensively studied through heavy-ion collisions (SPS, RHIC or LHC) and is still carefully analysed. The experiments seem to conclude that the QGP behaves like a perfect fluid with a low ratio viscosity over entropy around the critical temperature of deconfinement $T_c$. Therefore, this observation suggests that a strongly-coupled phase (called sQGP) is present in this temperature range and that the QCD confinement/deconfinement phase transition is much more driven by screening effects of the strong interaction. In order to correctly describe the different mechanisms at work during this phase transition, or more exactly during this crossover, a lot of theoretical researches (lattice QCD, phenomenological approaches...) are carried out. In particular, finding the QGP equations of state (EoS) is a crucial information nowadays needed. 

\par The aim of this work is to investigate the thermodynamic features of the QGP by resorting to a phenomenological approach based on $T$-matrix computations. This approach has the advantage to allow the study of bound and scattering states of the system in a whole picture. Relevant results have already been established for heavy-quarkonia above $T_c$ \cite{cabre06} and also for glueballs in the Yang-Mills plasma \cite{LSCB}. Moreover, in this latter reference, the EoS of the Yang-Mills plasma for SU(N) and G$_2$ have been computed thanks to the Dashen, Ma and Bernstein's formulation of statistical mechanics in terms of the ${S}$-matrix (or ${T}$-matrix)~\cite{dashen}.  Such a formulation is particularly well suited for systems whose microscopic constituents behave according to relativistic quantum mechanics. The QGP is indeed identified to a quantum gas of gluons, quarks and antiquarks, which are seen as the effective degrees of freedom propagating in the plasma. This assumption is actually common to all the so-called quasiparticle approaches \cite{quasip,quasip2}. However, thanks to the ${T}$-matrix  formulation, the strongly-interacting regime can also be investigated here, in which bound states are expected to still survive above $T_c$ \cite{Shur04}. 

\par The paper is organized as follows. Sec.~\ref{Tmatsec} is a summary of the approach used here and about which detailed explanations can be found in \cite{LSCB, LSB}. In Sec.~\ref{param}, the model parameters are presented and discussed. In particular, the quasiparticle bare masses are extracted from the $T = 0$ spectrum. In Sec.~\ref{BS_QGP}, the binary bound state spectrum above $T_c$ is computed and analysed. Then, the EoS of the deconfined phase at zero baryonic potential are studied for $N_f = 2$ and $N_f = 2 + 1$ in Sec.~\ref{EoS_QGP}. To finish, an exploratory work at small baryonic potential is carried out in Sec.~\ref{EoS_QGP_mu}. All our EoS are compared to recent lattice QCD (lQCD) ones. Our results are finally summarized in Sec.~\ref{conclu}.

\section{$\bm T$-matrix formalism in statistical physics}\label{Tmatsec}

\subsection{Generalities}\label{gen}

\par The results of Dashen, Ma and Bernstein \cite{dashen} establishing the grand potential of an interacting relativistic particle gas $\Omega$, expressed as an energy density, is given by (in units where $\hbar=c=k_B=1$)
\begin{equation}\label{pot0}
\Omega=\Omega_0+\sum_\nu\left[\Omega_\nu-\frac{{\rm e}^{\beta\vec \mu\cdot\vec  N}}{2\pi^2\beta^2}\int^\infty_{M_\nu} \frac{dE}{4\pi i}\, E^2\,  K_2(\beta E)\, \left. {\rm Tr}_\nu \left({\cal S}S^{-1}\overleftrightarrow{\partial_E}S \right)\right|_c\right]\text{.}
\end{equation}  
This equation is made of two parts. The first term $\Omega_0$ refers to the grand canonical potential of the free relativistic (quasi)particles, while the second term accounts for interactions in the plasma. This latter is made of a sum running on all the species, the number of particles included, and the quantum numbers necessary to fix a channel. The vectors $\vec \mu=(\mu_1, \mu_2,\dots)$ and $\vec N=(N_1,N_2,\dots)$ contain the chemical potentials and the particle number of each species taking part in a given scattering process. The set of all these channels is generically denoted $\nu$. As usual, the chemical potential $\mu$ is the Lagrange multiplier associated with the number of particles. It is a measure for the density of particles. In relativistic models, the sign of $\mu$ is changed, passing from matter to antimatter. This parameter marks imbalance between matter and antimatter \cite{dashen, Kapu06}.

\par One can notice that the contribution of the bound and scattering states are decoupled. The threshold $M_\nu$ is the summation on the masses of all the particles included in a given channel $\nu$. Below $M_\nu$, bound states appearing as pole in the $S$-matrix (equivalently $T$-matrix) are added as free additional species: $\Omega_\nu$ is the grand canonical potential describing a free relativistic gas of the $\nu$-channel bound states. Above $M_\nu$, the scattering contribution is expressed as an integration depending on a trace, taken in the center-of-mass frame of the particles in the channel $\nu$, and function of the $S$-matrix of the system. $S$ is in particular a function of the total energy $E$. The symmetrizer ${\cal S}$ enforces the Pauli principle when a channel involving identical particles is considered, and the subscript $c$ means that only the connected scattering diagrams are taken into account. $K_2(x)$ is the modified Bessel function of the second kind and $\beta = 1/T$ where $T$ is the temperature. The symbol $A\overleftrightarrow{\partial_x} B$ denotes $A(\partial_xB)-(\partial_xA)B$.

\par By definition, $S$ is linked to off-shell $T$-matrix  ${\cal T}$:
\begin{equation}
S=1-2\pi i\, \delta(E - H_0)\, {\cal T}\text{,}
\end{equation} where $H_0$ is the free Hamiltonian of the system. As in \cite{LSCB, LSB}, we will only focus on two-body channels. So, a way to obtain ${\cal T}$ is to solve the Lippmann-Schwinger equation, schematically given by
\begin{equation}\label{ls}
{\cal T}=V+ V\, G_0\, {\cal T} \text{,}
\end{equation}
with $G_0$ the free two-body propagator and $V$ the interaction potential. It is worth mentioning that for three-body channels, Faddeev equations should be used  in order to eliminate the spurious solution of the Lippmann-Schwinger equation \cite{joachain}. Such considerations will be thus out of scope in this paper. 

\par Once Eq.~(\ref{pot0}) is computed, all thermodynamic observables can derived. For example, the pressure is simply given by 
\begin{equation}
p=-\Omega .
\end{equation}

The sum $\sum_\nu$ appearing in (\ref{pot0}) explicitly reads $\sum_{I}\sum_{J^{PC}}\sum_{{\cal C}}$, where only two particles are involved in the interaction process, $I$ is a possible isospin channel, ${\cal C}$ is the color channel, and $J^{PC}$ is the spin/helicity channel (the labels $C$ or $P$ must be dropped off if the charge conjugation or the parity are not defined). 

The normalized trace anomaly can also be computed by the following formula
\begin{equation} \label{ta}
\frac{\Delta}{p_{SB}} =  -\beta \left( \partial_\beta \frac{p}{p_{SB}} \right)_{\beta\mu}.
\end{equation}
\noindent where $p_{SB}$ is the Stefan-Boltzmann pressure. Although we give here some results about the trace anomaly, it is mentioned in \cite{LSCB} that some improvements must be done in order to obtain a fully reliable estimation of this quantity. 


\subsection{Quasiparticle properties} \label{properties}

\par Assuming that the dominant scattering processes are the two-body ones, a key ingredient of the present approach is the two-body potential $V$, encoding the interactions between the particles in the plasma. As in \cite{LSCB, LSB}, $V$ is extracted  from the static quenched SU(3) lQCD free energy $F_1$, between a $q\bar{q}$ pair in singlet representation at finite temperature \cite{lPot}, and then fitted with a Cornell potential, screened thanks to the Debye-H\"uckel theory \cite{Satz90}(see Appendix~B in \cite{LSCB}). Note that unquenched lQCD results are also available in \cite{Kacz}. Nevertheless, since these results are not significantly different from the quenched ones, the quenched potential is kept as basis of our computations, giving the accuracy expected in our work. 
\par From that, the internal energy $U_1$ is computed, $U_1 = F_1 - T \partial_T F_1$, and considered as the interaction potential. This choice is still a matter of debate. Nevertheless, it has given correct results in the ordinary YM case, as shown in \cite{LSCB}. Moreover in Sec~\ref{QGP_3}, we will see that, according to our prescription for the quasiparticle masses, the internal energy is required to have a better agreement between our results and lQCD ones just above $T_c$. 

\par No relativistic corrections will be taken into account for light-quark interactions within this paper. Indeed, the quasiparticle quark masses used in our approach are large enough to assume static potentials at first approximation. Nevertheless, this task is left for future works. 

\par Moreover, all hyperfine interactions are neglected. We can expect that they are non-dominant with respect to the spin-independent contributions, since these processes are assumed to depend on the inverse square of the effective mass. With this hypothesis, we also miss the diagonal annihilation contributions.

Finally, the Casimir scaling is used to extract the leading-order gauge dependence of $U_1(r,T)$ for $T > T_c$, as proposed in Sec.~II in \cite{LSCB}. The Casimir scaling means that potentials between colored sources are proportional to the eigenvalues of the quadratic Casimir operator for their representation \cite{bali}. It is the simplest color dependence for two color sources: It has indeed the same form as the one for the one-gluon exchange process. Nevertheless, it is important to stress that the interaction considered within this paper contains other processes since it stems from a lQCD computation. Let us note that the annihilation mechanism, which does not respect the Casimir scaling, is a contact interaction and is then vanishing for all non-S states. Moreover, it is worth mentioning that the Casimir scaling seems very well respected between two static color sources in the $T=0$ sector \cite{bali}. Computations in the $T>T_c$ sector show a situation which is slightly different: The Casimir scaling seems partly violated (at most 20\%) for short distances and temperatures near $T_c$ \cite{gupta}. Nevertheless in this work, as in \cite{LSCB}, we assume that the Casimir scaling is satisfied. The final form of the potential is thus the following one:
\begin{equation}
V(r,T) = \displaystyle\frac{\kappa_{{\mathcal{C};ij}}}{\kappa_{{\bullet; q\bar{q}}}} \left[U_1(r,T) - U_1(\infty , T) \right], 
\label{Used_pot}
\end{equation}
\noindent where 
\begin{equation}
\kappa_{{\mathcal{C};ij}} = \displaystyle\frac{C_2^{\mathcal{C}} - C_2^{R_i} - C_2^{R_j}}{2 C_{2}^{\text{adj}}},
\label{color_scaling}
\end{equation}
and where $C_2^R$ is the quadratic Casimir of the representation $R$. $\mathcal{C}$, adj, $R_i$ and $R_j$  stand respectively for the pair, adjoint, $i$- and $j$-particle representation. For instance, 
\begin{equation}
C_{2}^{\text{adj}} = N \text{, } C_2^{\bullet} = 0 \text{, } C_2^{q} = C_2^{\bar{q}} = \displaystyle\frac{N^2 - 1}{2N} ,
\end{equation}
for the SU(N) gauge group (the singlet representation is denoted by $\bullet$). All the values taken by \eqref{color_scaling} for the various color channels considered in this study are given in Appendix~A in \cite{LSCB}. Let us note that the interaction can be attractive or repulsive. The normalization of \eqref{Used_pot} is given by $\kappa_{{\bullet; q\bar{q}}}=-4/9$,  since $U_1(r,T)$ is fitted on a singlet $q\bar{q}$ potential for SU(3). We can also notice in \eqref{Used_pot} that the long-distance behavior of the lattice potential $U_1(\infty, T)$, is subtracted. Indeed, this term is assimilated, as suggested in \cite{Mocs06}, as a thermal mass contribution for the quasiparticles. Moreover,  it ensures the convergence of the scattering equation and the possibility to perform the Fourier transform.

\par When the quasiparticles are infinitely separated, the only remaining potential energy can be seen as a manifestation of the in-medium self-energy effects, $U_1(\infty, T) = 2m_q(T)$. We thus encode these effects as a mass shift $\delta(T)$ to the ``bare" quasiparticle mass $m_0$, by following the arguments exposed in \cite{LSCB}:
\begin{equation}
m(T)^2 = m_{0}^2 + \delta(T)^2 \text{.}
\label{mg}
\end{equation}
\noindent In order to get the thermal mass for any particles, the first-order color dependence is extracted in agreement with the hard-thermal-loop (HTL) leading-order behavior \cite{Blai99}: 
\begin{equation} \label{delta}
\delta(T) = \sqrt{\frac{C_2^R}{C_2^{\text{adj}}}} \Delta (T) ,
\end{equation}
where the quantity $\Delta (T)$ is assumed to be color-independent. As $U_1(r,T)$ is fitted on a singlet $q\bar{q}$ potential for $SU(3)$, we have here
\begin{equation}
\displaystyle\frac{U_1(\infty)}{2} = m_q(T) = \sqrt{\frac{C_2^q}{C_2^{\text{adj}}}} \Delta (T) = \frac{2}{3} \Delta (T).
\end{equation}
\noindent For further details about the behavior of $m(T)$, one can refer to Sec.~V in \cite{LSCB} and to Sec.~\ref{param} in this paper. At this stage, one has to have in mind that chiral symmetry is not taken into account in our formalism. Comments about that issue will be given in the conclusions.


\subsection{Solving Lippman-Schwinger equations}\label{LSeq}

\par The Lippman-Schwinger equation leading to the on-shell $T$-matrix can be computed from \eqref{ls} as in \cite{LSCB, LSB}: 
\begin{eqnarray}\label{tosolve}
{\cal T}_\nu(E; q,q') &=& V_\nu(q,q') + \frac{1}{8\pi^3} \int_0^\infty dk\, k^2\, V_\nu(q,k)  \\
&& \times \,G_0(E;k)\, {\cal T}_\nu(E;k,q') \left[1 \pm f_{p1}(\epsilon_1)\right] \left[1 \pm f_{p2}(\epsilon_2)\right], \nonumber
\end{eqnarray}
where $E$ is the energy in the center-of-mass frame, $\epsilon_i$ the asymptotic energy of the particle $i$, and where the free two-body propagator is computed thanks to the Blanckenbecler-Sugar (BbS) reduction scheme. Its explicit form is given in Appendix~C in \cite{LSB}. Moreover,  the in-medium effects, namely the Bose-enhancement and the Pauli-blocking are included following  \cite{Prat94}. $f_p$ is thus the distribution function of the $p$-species: 
\begin{equation}
f_p (\epsilon) = \frac{1}{e^{\beta(\epsilon - \mu)} \mp 1} ,
\end{equation}
the $-$ stands for bosons while the $+$ for fermions, and $\mu$ is a possible chemical potential. The sign choice in \eqref{tosolve} also depends on the nature of the particles: $+$ for bosons and $-$ for fermions. Let us note that the impact of these in-medium effects on our EoS is very small. Therefore, the results obtained in \cite{LSCB} remain valid. 

\par Concerning the interaction potential $V_\nu(q,q')$ entering in~\eqref{tosolve}, it is obtained by the Fourier transform of the interaction extracted in lQCD. Since our potential has a spherical symmetry, we have
\begin{equation}\label{Vqq}
V(q,q',\theta_{q,q'}) = 4 \pi \displaystyle\int_0^\infty dr \, r V(r) \displaystyle\frac{\sin(Q\,r)}{Q} ,
\end{equation}
where $Q = \sqrt{q^2 + q'^2 - 2 q q' \cos \theta_{q,q'}}$ and $\theta_{q,q'}$ is the angle between the momenta $\vec q$ and $\vec q\,'$.
\par For channels given by ordinary $\left|^{2S+1}L_J\right\rangle$ states, $V_\nu(q,q')$ is obtained from
\begin{equation}
V_L(q,q') = 2\pi \displaystyle\int_{-1}^{+1} dx P_L(x) V(q,q',x),
\end{equation}
where $P_L$ is the Legendre polynomial of order $L$. The spin $S$ is not indicated since our interaction is spin-independent.
\par When at least one particle is transverse, the helicity formalism \cite{jaco} has to be used. It is then very convenient to decompose a helicity state in the basis states $\left|^{2S+1}L_J\right\rangle$ in order to perform the computations. For a particular helicity state $\left|J^P\right\rangle$, it reads 
\begin{equation}\label{helexpspin}
\left|J^P\right\rangle = \sum_{L,S} C_{L,S} \left|^{2S+1}L_J\right\rangle. 
\end{equation}
Then, it can be shown that 
\begin{equation}
V_{J^P}(q,q') = \sum_{L,S} C_{L,S}^2 V_L(q,q'),
\end{equation}
since our interaction is spin-independent. All the helicity states needed for this study are listed in Appendix~B in \cite{LSB}.

\par The Haftel-Tabakin algorithm is a reliable procedure to solve the $T$-matrix problem \cite{cabre06,haftel}. The momentum integral is discretized within an appropriate quadrature, thus turning the integral equation into a matrix equation, namely $\sum {\cal F}_{ik} {\cal
T}_{kj} = V_{ij}$, where schematically, 
\begin{equation}
{\cal F}=1-wVG (1 \pm f_{p_1}) (1  \pm f_{p_2})
\end{equation}
and where $w$ denotes the integration weight. The solution follows trivially by matrix inversion. 
Bound states are naturally poles below $M_\nu$. An interesting criterion for finding them is to use the determinant of the transition function $\cal F$ (referred to as the Fredholm determinant) since it vanishes at the bound state energies \cite{LSCB}. Finally, once ${\cal T} (E;q,q')$ is known, the on-shell $T$-matrix is readily obtained
as ${\cal T}(E;q(E),q(E))$, with $q(E)$ given by 
\begin{equation}
q(E) = \frac{\sqrt{\left( E^2 - (m_{1} + m_{2})^2 \right) \left( E^2 - (m_{1} - m_{2})^2 \right)}}{2E}\text{.}
\label{diff_mass}
\end{equation}


\section{Model parameters}\label{param}

\subsection{Assumptions}
\par Before fixing the parameters and applying the general formalism described in the previous section to the $N_f = 2~(+1)$ QGP, let us discuss some general assumptions that we have done within this model.
\par In our approach, there are different species of quasiparticles: the gluons ($g$), the light (anti)quarks ($l$, resp. $\bar{l}$), the strange (anti)quarks ($s$, resp. $\bar{s}$) and the heavy (anti)quarks ($c$ and $b$, resp. $\bar{c}$ and $\bar{b}$). The (anti)quarks are spin-1/2 particles belonging to fundamental (conjugate) representation of the gauge group. Despite their nonvanishing mass, the gluons are transverse spin-1 bosons in the adjoint representation. The gluon mass is dynamically generated by self-energy effects, which does not imply a drastic change of their nature. It has been shown that the gluon must be considered as a transverse spin-1 boson to reproduce correctly the expected glueball spectra at $T = 0$ \cite{Math08}. Moreover, some lattice data support the presence of massive transverse modes only in a gluon plasma \cite{quasip}.  The two-body channels to be considered are $gg$, $qq$, $\bar{q}\bar{q}$, $q\bar{q}$, $qg$ and $\bar{q}g$. The lowest corresponding spin/helicity states are given in Appendix~B of \cite{LSB}, and the possible color channels can be found in Appendix~A of \cite{LSCB}. Within this study, we only focus on the SU(3) gauge group.

\par Let us examine the different possibilities of interactions: 
\begin{itemize} 
\item As explained above, the interaction (\ref{Used_pot}) between two gluons or two quarks follows strictly the Casimir scaling and neglects all hyperfine corrections, annihilation ones included. 
\item Although this interaction is expected to take into account complicated exchanges (since it stems from a lQCD calculation), it is interesting to look at the simplest possible Feynman diagrams between two particles. Two gluons or two quarks can exchange a gluon, but the basic gluon-quark interaction is a quark exchange. So the choice (\ref{color_scaling}) for the color factor is questionable for this particular interaction. To correct this point is beyond the scope of this work, but it is worth mentioning that the contributions of the $qg$ and $\bar{q}g$ interactions is expected to be very weak in our model (see Sec.~\ref{EoS_QGP}). 
\item Processes transforming a $gg$ pair into a $q \bar{q}$ pair exist, but we have checked that mechanisms of order 1 are naturally suppressed since there is no overlap between $gg$ and $q \bar{q}$ states \cite{LSB}. As we neglect second order processes, as hyperfine interactions, we do not take into account transition between $gg$ and $q \bar{q}$ pairs.
\end{itemize}

\subsection{Potential at $\bm{T = 0}$} \label{T0BS}

\par In order to fix our parameters for starting the computations at finite temperature, and to check the validity of our model, some pieces of information can be extracted from the $T = 0$ bound state spectrum as in \cite{LSCB}. 

\par In quenched SU(3) lattice QCD, the potential between a static quark-antiquark pair at zero temperature is compatible with the funnel form 
\begin{equation}
V_f(r) = \sigma r -\frac{4}{3}\frac{\alpha}{r}, 
\end{equation}
where $\alpha = 0.4$  and $\sigma = 0.176$~GeV$^2$ (standard values for the running coupling constant $\alpha$ and the string tension $\sigma$ at $T=0$). Again, we neglect the contributions of annihilation processes. Since the Fourier transform of $V_f(r)$ is not defined (because of a non-zero asymptotic value), a string-breaking value $V_{sb}$, has to be introduced in order to make it convergent \cite{cabre06}. $V_{sb}$ is thus seen as the energy above which a light quark-antiquark pair can be created from the vacuum and breaks the QCD string. This scale is then subtracted and the potential effectively taken into account is $V_f(r)-V_{sb}$, while $V_{sb}/2$ is interpreted as an effective quark mass using the same arguments as those detailed in Sec.~\ref{properties}. 

\par According to the color scaling \eqref{color_scaling}, the potential describing the interactions between two color sources (with representations $R$ and $\bar R$) at zero temperature, is  
\begin{equation}\label{T0pot}
V_0(r) = \frac{9}{4}\left( C_2^{R} + C_2^{\bar R} \right) V_f(r) - V_{sb}^{R\bar{R}} \text{,}
\end{equation}
\noindent since $C_2^{\bullet} = 0$. The factor $9/4$ appears since the potential $V_f$ is fitted on a singlet $q\bar{q}$ pair for a SU(3) gauge group. In this case, $V_{sb}^{R\bar{R}}$ should rather be interpreted as the energy scale necessary to form two sources of color compatible with the existence of the two new color singlet pairs of particles created by the string breaking. As in \cite{LSCB, LSB}, if $m_0$ is the bare mass of the particle, the $T=0$ mass $m(0)$, used to compute the bound state is then
\begin{equation}
m(0)^2 = m_0^2 + \left(\displaystyle\frac{V_{sb}^{R\bar{R}}}{2}\right)^ 2,
\end{equation}
keeping the same structure as in \eqref{mg}.

\subsection{$\bm{T = 0}$ Bound State Spectrum} \label{param_T0}

\par The zero-temperature spectrum of the theory can be computed by solving \eqref{tosolve} with the potential~\eqref{T0pot}. As mentioned in Sec.~\ref{LSeq}, instead of looking at the pole of the $T$-matrix, the zeros of $\det {\cal F}$ are computed in order to establish the bound state spectrum. 

\par The lightest glueball spectrum, namely the $0^{++}$, $0^{-+}$ and $2^{++}$, has already been computed in \cite{LSCB}. The parameters, $V_{sb}^{gg}$ and $m_{0}$, were respectively fixed to 2~GeV and 0.7 GeV.  $V_{sb}^{gg}$ is in agreement with lattice data showing that the mass of the lightest gluelump is given by $0.85(17)$~GeV \cite{Bali04}, while $m_{0}$ is an acceptable value for the zero-momentum limit of the gluon propagator at zero temperature in view of previous studies locating this mass typically between 500 and 700 MeV (see \textit{e.g.} \cite{Corn82,Oliv11,Bino09}). Interesting reader can refer to \cite{LSCB} for additional information about the $T = 0$ glueball spectrum.

\par Within this paper, the stress is put on mesons with an orbital angular momentum $L = 0$ or $L = 1$. The allowed states with these quantum numbers are displayed in Table~\ref{Tab_qqbar}. As it can be seen, several $J^{PC}$ states are associated to a same $L$.  Since the potential \eqref{T0pot} does not depend on other quantum numbers, all these states are degenerate within our approach. 
\begin{table}[h!]
\begin{center}
\begin{tabular}{|c|c|c|c|}
\hline 
$J$ & $L$ & $S$ &   $J^{PC}$ \\ \hline 
0 & 0 & 0 & $0^{-+}$\\
& 1 & 1  & $0^{++}$ \\ \hline
1 & 1 & 0  & $1^{+-}$ \\
& 0 & 1 &  $1^{--}$ \\
& 1 & 1 &  $1^{++}$\\ \hline
2 & 1 & 1 &  $2^{++}$ \\ \hline
\end{tabular}
\caption{\label{Tab_qqbar} $J^{PC}$ states allowed for $q\bar{q}$ with $L = 0$ or $L = 1$ at $T = 0$. The parity of the state $P$, is given by $(-1)^{L+1}$ while the charge conjugaison $C$, is $(-1)^{L+S}$. }
\end{center}
\end{table}

\par The used parameters are summarized in Table~\ref{Tab_param_T0}. There are essentially two main points to notice. Firstly, a shift of 0.3 GeV to the PDG quark bare mass \cite{PDG} is systematically present. It is a common assumption within quasiparticle approaches since this shift corresponds to one third of the nucleon mass. Moreover, it is a typical value for the chiral condensate according to \cite{Bowm}. Secondly, the string breaking depends on the quark flavor. This could be explained by the following argument: It is not the same region of the potential that is relevant for the dynamics of all the quark flavors. Indeed, light quarks are more sensitive to the linear part of the interaction while the heavy-quark potential is dominated by the Coulomb one. 
\par According to this interpretation, $V_{sb}^{q\bar{q}}$ has to be higher for light quarks and has to decrease with heavier quark flavors. It is exactly what it is observed with the parameters in Table~\ref{Tab_param_T0}. Moreover, when the bound state is made of two different quark flavors, $V_{sb}^{q\bar{q}}$ takes a value between the chosen string breaking for the two associated quarkonia systems. For $D$ and $B$-mesons, $V_{sb}^{q\bar{q}}$ is closer to the string breaking of heavy quarkonia.

\begin{table}[h!]
\begin{center}
\begin{tabular}{|c|ccc|}
\hline
Quark composition & $V_{sb}^{q\bar{q}}$ & $m_0^1$ & $m_0^2$ \rule[-7pt]{0pt}{20pt} \\ \hline
Light (l-l) & 2.6 & 0.3  & 0.3\\ 
Strange (s-s) & 2 & 0.4 & 0.4 \\
Charm (c-c)& 1 & 1.6 & 1.6\\
Beauty (b-b) & 0.7 & 4.95 & 4.95 \\
Kaon (l-s) & 2.4 & 0.3 & 0.4 \\ 
D-meson (l-c) & 1.5 & 0.3 & 1.6 \\
D$_s$-meson (s-c) & 1.2 & 0.4 & 1.6 \\
B-meson (l-b) & 1.2 & 0.3 & 4.95  \\
B$_s$-meson (s-b) & 1 & 0.4 & 4.95 \\
B$_c$-meson (c-b) & 0.7 & 1.6 & 4.95 \\ \hline
\end{tabular}
\caption{\label{Tab_param_T0} Masses and string breaking (in GeV) for the different flavors of quarks.} 
\end{center}
\end{table}

\par In Table~\ref{TabMesonExp}, the results are compared to experimental \cite{PDG}. As it can be noticed, a quite good agreement is reached provided that we do not consider the lightest pions and kaons \textit{i.e.} $\pi (140)$ and $K(495)$. Indeed, the fact that the mass of these lightest mesons are not achievable can be explained by the theoretical origin of such states: The pion is the Goldstone boson resulting from the spontaneously breaking of the chiral symmetry. A so peculiar phenomenon can not be described within such simple effective model. Moreover, according to quasiparticle standard approaches, the spin effects are the weakest in a $S = 1$ channel. Since our computations do not take into account such effects, it is reasonable that our results for the $L = 0$ light mesons are closed to the $\rho$ instead of the $\pi$.

\begin{table}[h!]
\begin{center}
\begin{tabular}{|c|cc|c|cc|}
\hline
$L = 0$ & Exp.& $T$-matrix & $L = 1$ &  Exp.& $T$-matrix \\ \hline
$\rho (u\bar{u}, d \bar{d})$ & 0.77 & 0.72 & $a_0 (u\bar{u}, d \bar{d})$ & 1.45 & 1.45\\
$\Phi (s\bar{s})$ & 1.02 &  1.08 & $f_2'(s\bar{s})$ & 1.53 & 1.58\\
$K^*(l$-$s)$ & 0.89 & 0.89 & $K (l$-$s)$ & 1.43 & 1.52 \\
$J/\psi (c\bar{c})$ & 3.10 & 3.01 & & & \\
$\Upsilon (b\bar{b})$ & 9.46 & 9.40 &  & & \\
$D^* (l$-$c)$ & 2.01 & 2.01 &  &  &  \\
$D_s^* (s$-$c)$ & 2.11 & 2.11 &  &  &  \\
$B^* (l$-$b)$ & 5.33 & 5.33 &  &  &  \\
$B_s^* (s$-$b)$ & 5.42 & 5.41 &  &  &  \\
$B_c^* (s$-$c)$ & - & 6.39 &  &  &  \\
\hline
\end{tabular}
\caption{\label{TabMesonExp} Masses (in GeV) of the $L = 0$ and $L = 1$ meson states  at zero temperature with the gauge group SU(3). Our results (third and sixth column), are compared to the experimental data of \cite{PDG} (second and fifth column).} 
\end{center}
\end{table}
\par Finally, let us add that, unlike in the glueball case, the $T = 0$ meson mass depends on the gauge group since $\kappa_{\bullet ;q\bar{q}}$ depends on it (see Appendix~A in \cite{LSCB}). Within our approach, such study is not difficult to carry out. The meson mass dependence in function of the gauge group is not studied here, since the principal interest of the $T$-matrix computations at $T = 0$ is to extract and to check the parameters  we will use at $T \neq 0$. In this regard, let us note that only the quasiparticle bare masses will enter in our computations at $T \neq 0$ and not the string breaking $V_{sb}$.  What can be said however about the gauge-group dependence is that, in the case of SU($N$), the meson masses are of order 1 as expected, see conclusions for more comments about this.

\subsection{Critical temperature of deconfinement} \label{param_Tc}

\par The last global parameter that has to be fixed within our approach is the critical temperature of deconfinement $T_c$. As we fit our interactions on lattice calculations, the definition of $T_c$ comes from these approaches: the color averaged as well as the singlet free energy of a quark-antiquark pair will tend towards finite nonzero values for all temperatures $T > T_c$ (and diverge below $T_c$). In \cite{LSCB}, a value of 0.3 GeV was used since the focus was only on the gluon sector. Here, in order to stay coherent with current lattice data, $T_c$ is moved to 0.15 GeV. This change naturally modifies the thermodynamics in the gluon sector that was established in \cite{LSCB}. Let us discuss this point. 

\par First of all, let us set $z = T/T_c$. The two-body interaction potential between particles only depends on $z$ as it can be explicitly shown from its expression given in Appendix~B of \cite{LSCB}. Therefore, it is the same for the gluon thermal mass $\delta_g (T)$,  according to \eqref{delta}. Moreover, since the effect of the Bose-enhancement can be considered as negligible, it can also be assumed that the $T$-matrix $T_{J^{PC}}$, is only a function of $z$. 

\par
\begin{figure}[h!]
\begin{center}
\includegraphics*[width=0.49\textwidth]{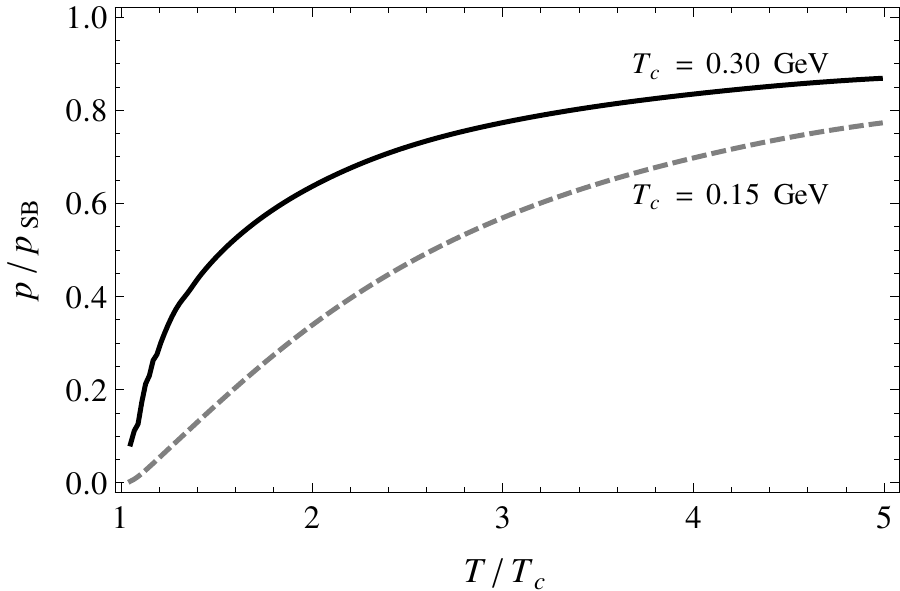}
\includegraphics*[width=0.49\textwidth]{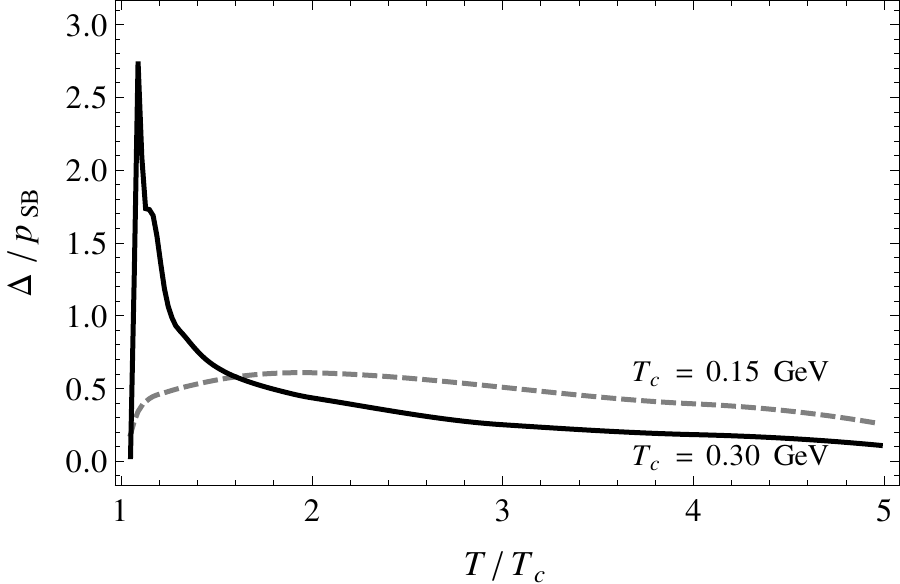}
\caption{Normalized pure-gauge pressure $p/p_{SB}$, and trace anomaly $\Delta/p_{SB}$ (without bound states) versus temperature in units of $T_c$ (with $T_c = 0.15$ and 0.3~GeV).}
\label{CompareTemp}
\end{center}
\end{figure}

 In Fig.~\ref{CompareTemp}, the $T_c$-impact is analyzed for the pure gauge EoS with $T_c =$ 0.15 and 0.30 GeV, all other parameters remaining fixed. The way to compute these EoS is given in \cite{LSCB} and will be recalled in Sec.~\ref{Gen_EoS}. As it can be noticed, the general behavior is not the same and the normalized pressure seems to increase when $T_c$ increases. It is especially worth remarking that, if we only consider the contribution of the quasiparticle ideal gas, $T_c$ has an impact on the thermodynamics. Indeed, the pressure depends on the ratio $m/T = m(z)/(zT_c)$, depending on $T_c$ for a fixed $z$. 
\par
\begin{figure}[h!]
\begin{center}
\includegraphics*[width=0.49\textwidth]{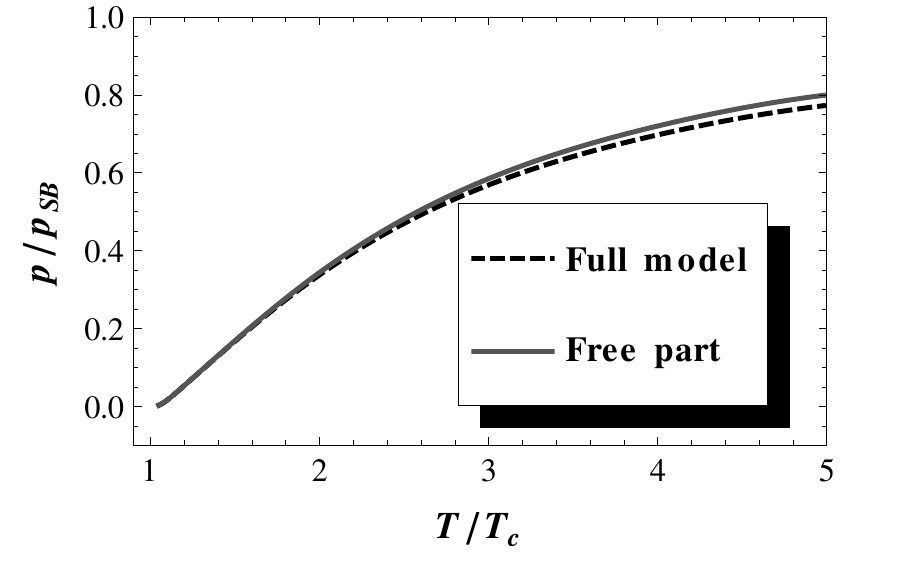}
\includegraphics*[width=0.49\textwidth]{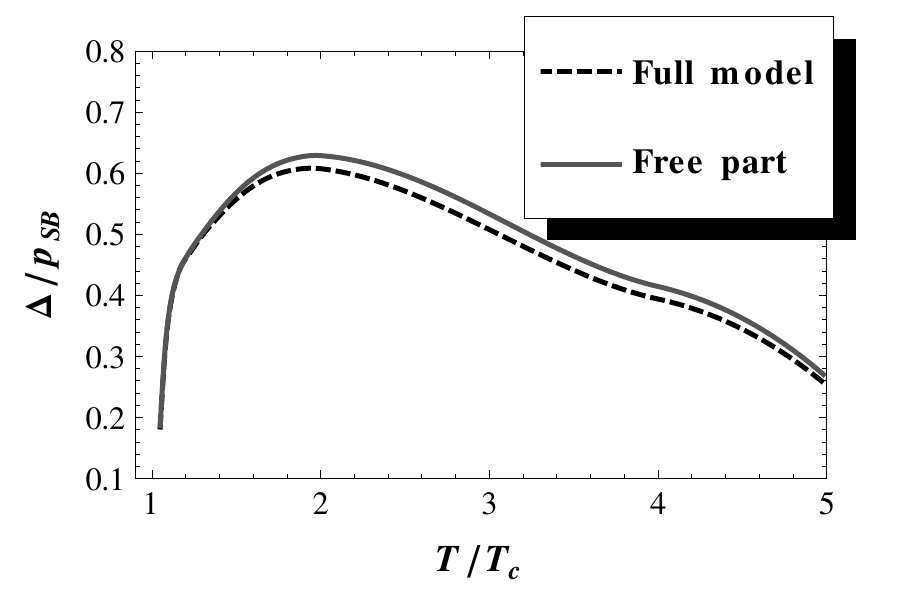}
\caption{(Left) Normalized pure-gauge pressure $p/p_{SB}$ versus temperature in units of $T_c$ (with $T_c = 0.15$~GeV), compared to the free part contribution. (Right) Normalized pure-gauge trace anomaly $\Delta/p_{SB}$ versus temperature in units of $T_c$ (with $T_c = 0.15$~GeV), compared to the free part contribution.}
\label{CompareTemp150}
\end{center}
\end{figure}

 The behavior of the normalized trace anomaly (without bound states) is also presented in Fig.~\ref{CompareTemp}. The $T_c$-dependence is not very easy to predict because the trace anomaly depends on the slope of the associated pressure curve, and small variations can generate a drastic change. In Fig.~\ref{CompareTemp}, we can indeed observe that the behavior around $T_c$ is extremely different for $T_c = 0.15$ and 0.3 GeV. The peak structure is lost for $T_c = 0.15$ GeV. This could be due to two reasons: the total change of the free part structure and the small impact of the interactions in comparison with the EoS obtained in \cite{LSCB}, as it is shown in Fig.~\ref{CompareTemp150}.


\section{Bound States within the QGP} \label{BS_QGP}

\par The existence or not of bound states in the deconfined phase is not forbidden in principle, especially around $T_c$ where interactions are expected strong enough to bind two or more particles \cite{Shur04}. Since the operator $\kappa_{\mathcal{C};ij}$ is negative for several color channels, the finite-temperature spectrum of QCD above $T_c$ can be computed by solving \eqref{tosolve} with the potential~\eqref{Used_pot}. The thermal masses of the quasiparticles are given by \eqref{mg} with $m_0$ extracted from the $T = 0$ bound state spectrum (see Sec.~\ref{param_T0}). 
\par Within our formalism, the channels in which bound states are favored at most should contain a $S$-wave component to avoid the centrifugal barrier and should have a symmetry that allows the state to be in a color singlet, the color channel in which the interactions are maximally attractive. 
 \begin{table}[h!]
\begin{center}
\begin{tabular}{|c|c|c|c|}
\hline
\multicolumn{4}{|c|}{Light quark sector} \\ \hline
\multicolumn{2}{|c|}{Channel} & $q\bar{q}~(L = 0)$ & $qq/\bar{q}\bar{q}~(L = 0)$ \\  
\multicolumn{2}{|c|}{${\cal C}$} & Singlet & AS  \\  \hline
$T/T_c$ & $2\,m_l$ && \\  
1.05 & 1.67 & 1.51 & 1.67 \\ 
1.10 & 1.28 &-& -\\ \hline
\end{tabular}
\begin{tabular}{|c|c|c|c|}
\hline
\multicolumn{4}{|c|}{Light-strange quark sector} \\ \hline
\multicolumn{2}{|c|}{Channel} & $q\bar{q}~(L = 0)$ & $qq/\bar{q}\bar{q}~(L = 0)$ \\  
\multicolumn{2}{|c|}{${\cal C}$} & Singlet & AS \\  \hline
$T/T_c$ & $m_l + m_s$ & &  \\  
1.05 & 1.71 & 1.54 & 1.71 \\ 
1.10 & 1.43 &-&- \\ \hline
\end{tabular}
\begin{tabular}{|c|c|c|c|}
\hline
\multicolumn{4}{|c|}{Strange quark sector} \\ \hline
\multicolumn{2}{|c|}{Channel} & $q\bar{q}~(L = 0)$ & $qq/\bar{q}\bar{q}~(L = 0)$ \\  
\multicolumn{2}{|c|}{${\cal C}$} & Singlet & AS \\  \hline
$T/T_c$ & $2\,m_s$ && \\  
1.05 & 1.76 & 1.57 &  1.74 \\ 
1.10 & 1.38 &1.38 & - \\
1.15 & 1.23 &- &  \\ \hline
\end{tabular}
\end{center}
\caption{\label{tab:all1} Masses (GeV) of lowest-lying QCD spectrum above $T_c$ ($T_c = 0.15$~GeV). Singlet and AS respectively refer to the singlet and antisymmetric representation of the gauge group. A line mark the temperature at which a bound state is not detected anymore. }
\end{table}
\par In the $gg$ case there are two such states: The $0^{++}$ and $2^{++}$ ones, in color singlet, correspond to the scalar and tensor glueballs respectively. We have observed in \cite{LSCB} that both the scalar and tensor glueball masses at $1.05$ $T_c$ were compatible with the zero-temperature ones. Moreover, the scalar glueball exists as a bound state up to 1.25 $T_c$ while the tensor one is bound up to 1.15 $T_c$. Note that in \cite{LSCB}, the impact of the Bose-enhancement  were not considered in the $T$-matrix. Therefore, the value of $T_c$ should modify the masses of the bound states. Nevertheless, it has been numerically checked that the data shown in \cite{LSCB} differ from the ones containing in-medium effects only with a relative error of the order of 2\%. This is the reason why they are not presented again here. 

\par Concerning the light and strange mesonic sector, the dissolution inside the plasma is much more rapid. Indeed, the most attractive channel, the $L = 0$ in singlet, is the only one that survives above $T_c$. Moreover, mesons quickly dissolve inside the plasma as it can be observed in Table~\ref{tab:all1}, and the meson masses are not compatible with the $T = 0$ one as in the $gg$ case.

\par The same assertion can also be drawn for the $qq$ and $\bar{q}\bar{q}$ sector. Indeed, since there is no more confinement (\text{i.e.}~only singlet representation) in the QGP, such states could exist above $T_c$; the most attractive one being the $L = 0$ in the antisymmetric representation (AS). Nevertheless, they also rapidly disappear just above $T_c$, as shown in Table~\ref{tab:all1}. Such rapid dissolution in comparison with the $gg$ case can be understood by the fact that the quark quasiparticle mass is lighter than the gluon one and that the $\kappa_{\bullet;gg}$ is more than two times the maximum magnitude of $\kappa_{{\cal C}}$ in the quark sector. Indeed,  $\kappa_{\bullet;gg} = -1$ while $\kappa_{\bullet;q\bar{q}} = -4/9$ and $\kappa_{AS;qq} = \kappa_{AS;\bar{q}\bar{q}} = -2/9$ for a SU(3) gauge group (see Appendix~A of \cite{LSCB}). Note that similar comments can also be done about the $qg$ and $\bar{q}g$ sector, leading to a quick melting of these bound states inside the plasma (see Table~\ref{qg_diss}). 
\begin{table}[h!]
\begin{center}
\begin{tabular}{|c|c|c|c|c|c|}
\hline
\multicolumn{6}{|c|}{\textbf{Charm quark sector}} \\ \hline
\multicolumn{2}{|c|}{Channel} & $q\bar{q}~(L = 0)$ & $qq/\bar{q}\bar{q}~(L = 0)$ & $q\bar{q}~(L = 1)$ & $qq/\bar{q}\bar{q}~(L = 1)$ \\  
\multicolumn{2}{|c|}{${\cal C}$} & Singlet & AS & Singlet & AS \\  \hline
$T/T_c$ & $2\,m_c$ && &&  \\  
1.05 & 3.56 & 3.14 &  3.44 & 3.48 & - \\ 
1.10 & 3.39 & 3.20 & 3.36  & - &\\
1.15 & 3.33 & 3.22 & -  &&\\
1.25 & 3.28 & 3.26 & &&\\
1.35 & 3.26 & - & && \\
\hline
\end{tabular}
\end{center}
\caption{\label{tab:all1} Masses (GeV) of lowest-lying charmonium states above $T_c$ ($T_c = 0.15$~GeV). Singlet and AS respectively refer to the singlet and antisymmetric representation of the gauge group. A line mark the temperature at which a bound state is not detected anymore.}
\label{quarkonia_c}
\end{table}

\par Concerning the heavy quark sector, quarkonia have already been studied within a similar $T$-matrix approach as the one proposed here \cite{cabre06}. The main differences are the inclusion of a relativistic correction to the potential in \cite{cabre06} and the way of implementing the quasiparticle masses. Within this paper, the procedure \eqref{mg} to determine the quasiparticle masses is applied and allows one to compute systematically a large panel of binary bound states made of different quark flavors. 
\begin{table}[h!]
\begin{center}
\begin{tabular}{|c|c|c|c|c|c|}
\hline
\multicolumn{6}{|c|}{\textbf{Beauty quark sector}} \\ \hline
\multicolumn{2}{|c|}{Channel} & $q\bar{q}~(L = 0)$ & $qq/\bar{q}\bar{q}~(L = 0)$ & $q\bar{q}~(L = 1)$ & $qq/\bar{q}\bar{q}~(L = 1)$ \\ 
\multicolumn{2}{|c|}{${\cal C}$} & Singlet & AS &  Singlet & AS \\  \hline
$T/T_c$ & $2\,m_b$ &&  && \\  
1.05 & 10.2 & 9.35 & 9.75 & 9.60 & 9.90\\ 
1.15 & 9.94 & 9.62 & 9.87 & 9.79 & 9.94\\ 
1.20 & 9.93 & 9.71 & 9.86 & 9.92 & -\\ 
1.25 & 9.93 & 9.70 & 9.90 & - & \\ 
1.30 & 9.92 & 9.75 & 9.90 && \\ 
1.50 & 9.92 & 9.79 & - && \\ 
2.00 & 9.91 & 9.88 & && \\
2.40 & 9.91 &  -  &  &&\\  \hline
\end{tabular}

\end{center}
\caption{\label{tab:all1} Masses (GeV) of lowest-lying bottonium states above $T_c$ ($T_c = 0.15$~GeV). Singlet and AS respectively refer to the singlet and antisymmetric representation of the gauge group. A line mark the temperature at which a bound state is not detected anymore.}
\label{quarkonia_b}
\end{table}

\par Our study for the heavy quarkonia is displayed  in Tables~\ref{quarkonia_c} and \ref{quarkonia_b}. Around $T_c$, the $J/\psi$ and $\Upsilon$ masses are compatible with the $T = 0$ spectrum, unlike for the light and strange mesons. Moreover, they significantly survive above $T_c$, even if the dissociation temperatures are lower than the ones found in \cite{cabre06}. $qq$ and $\bar{q}\bar{q}$ states can also be formed with the medium but they dissolve more rapidly than the associated quarkonia, due their weaker interaction potential.

\par An analyse for the $D$ and $B$ mesons has also been carried out as well as for the $qg$ and $\bar{q}g$ states for all the quark flavor considered here. The different temperatures of dissociation are displayed in Tables~\ref{meson_diss}, \ref{qq_diss} and \ref{qg_diss}. We can notice that the more the quasiparticles considered in the binary state are heavy, the more it survives significantly above $T_c$. 

\begin{table}[h!]
\begin{center}
\begin{tabular}{|c|c|c|c|c|}
\hline
\multicolumn{5}{|c|}{\textbf{Dissociation temperature: $q\bar{q}$}} \\ \hline
& Light & Strange & Charm & Beauty \\ \hline
Light & 1.10 $\pm$ 0.05 & 1.10 $\pm$ 0.05 & 1.15 $\pm$ 0.05 & 1.20 $\pm$ 0.05 \\ \hline 
Strange &  & 1.15 $\pm$ 0.05 & 1.15 $\pm$ 0.05 & 1.20  $\pm$ 0.05\\ \hline 
Charm & & & 1.35 $\pm$ 0.05 & 1.60 $\pm$ 0.1 \\ \hline
Beauty &  & &  & 2.4 $\pm$ 0.1 \\ \hline
\end{tabular}
\end{center}
\caption{Temperature of dissociation in units of $T_c$ for $L = 0$ mesons. }
\label{meson_diss}
\end{table}

\begin{table}[h!]
\begin{center}
\begin{tabular}{|c|c|c|c|c|}
\hline
\multicolumn{5}{|c|}{\textbf{Dissociation temperature: $qq$ and $\bar{q}\bar{q}$ }} \\ \hline
& Light & Strange & Charm & Beauty \\ \hline
Light & 1.10 $\pm$ 0.05 & 1.10 $\pm$ 0.05 & 1.10 $\pm$ 0.05 & 1.10 $\pm$ 0.05 \\ \hline 
Strange &  & 1.10 $\pm$ 0.05 & 1.10 $\pm$ 0.05 & 1.10 $\pm$ 0.05 \\ \hline 
Charm & &  & 1.15 $\pm$ 0.05 & 1.25 $\pm$ 0.05 \\ \hline
Beauty &  &&  & 1.50 $\pm$ 0.05\\ \hline
\end{tabular}
\end{center}
\caption{Temperature of dissociation in units of $T_c$ for $L = 0$  $qq$ and $\bar{q}\bar{q}$ states. }
\label{qq_diss}
\end{table}

\begin{table}[h!]
\begin{center}
\begin{tabular}{|c|c|c|c|}
\hline
\multicolumn{4}{|c|}{\textbf{Dissociation temperature: $qg$ and $\bar{q}g$}} \\ \hline
 Light & Strange & Charm & Beauty \\ \hline
 1.10 $\pm$ 0.05 & 1.10 $\pm$ 0.05 & 1.20 $\pm$ 0.05 & 1.25 $\pm$ 0.05\\ \hline 
 
\end{tabular}
\end{center}
\caption{Temperature of dissociation in units of $T_c$ for $qg$ and $\bar{q}g$ states . }
\label{qg_diss}
\end{table}
In \cite{Satz}, it is found that radially excited states are unlikely to survive above $T_c$ since they tend to melt below the phase transition because of string breaking effects at finite temperature. In our case, such states can be found but not in all channels. However, they quickly dissolve within the medium.


\section{Equation of state of the QGP at $\bm{\mu = 0}$} \label{EoS_QGP}

\subsection{General expression} \label{Gen_EoS}
\par Now that the bound-state sector is analyzed, it is possible to compute explicitly the EoS and so, to study the QGP thermodynamics. In what follows, the heavy quark states will be not included in our EoS. Indeed, their contributions to the bound-state and scattering parts of the grand canonical potential are expected to be small because of their large bound-state masses. 
\par Some preliminary lattice results about the influence of charm quarks on the EoS can be found in \cite{Baza}. It appears actually that charm quarks bring a significant contribution to the trace anomaly above $T_c$. However, a technical problem of the present approach is that discontinuity appear in the trace anomaly when bound states melt \cite{LSCB}. This problem is especially apparent when heavy quarks are involved. Hence, including heavy flavors in our computations would lead to results that are probably not reliable, and we prefer not to consider them.
\par Let us thus particularize \eqref{pot0} to a QGP with $N_f = 2~(+1)$. As in \cite{LSCB, LSB}, a two-body restriction is used: The considered interactions are  $gg$, $qq$, $\bar{q}\bar{q}$, $q\bar{q}$, $qg$ and $\bar{q}g$, in different colour and $J^{P(C)}$ channels. Therefore, the first term in (\ref{pot0}), \textit{i.e.}~the free relativistic gas is given by 
\begin{equation}
\Omega_0^{\text{QCD}}= \underbrace{2 \, {\dim }\, adj\, \omega_0^B(m_g, 0)}_{gluons} + \underbrace{2 \displaystyle\sum_{n = 1}^{N_f} {\dim }\, q_n\, \omega_{0}^F (m_{q_n}, \mu_{q_n})}_{quarks} + \underbrace{2 \displaystyle\sum_{n = 1}^{N_f} {\dim }\, \bar{q}_n\, \omega_{0}^F (m_{q_n}, -\mu_{q_n})}_{antiquarks} \text{,}
\end{equation}

\noindent where the gluons have a mass $m_g$, and the (anti)quarks $q_n$ ($\bar{q}_n$) a mass $m_{q_n}$, given by the prescription \eqref{mg} with the $m_0$ value extracted from the $T = 0$ spectrum. $\mu_{q_n}$ is the chemical potential of the considered quark flavor. They are set to zero within this section. So, no asymmetry between quarks and antiquarks is taken into account within the QGP. The particle degrees of freedom are the following. The gluon is a transverse spin-1 (so, two spin projections) boson lying in the adjoint representation of the gauge group, while the quark (resp.~antiquark), existing in $N_f$ different flavors, is a spin-1/2 fermion belonging in the fundamental (resp.~conjugate) gauge-group representation. The grand canonical potential per degree of freedom associated to a bosonic species $\omega_0^B(m, 0)$, and to a fermionic species $\omega_0^F(m, \mu)$, with mass $m$ are given by
\begin{eqnarray}
	\omega_0^B(m, 0) &=& \frac{1}{2\pi^2\beta}\int^\infty_0dk\, k^2 \ln\left(1-{\rm
e}^{-\beta\sqrt{k^2+m^2}}\right)\text{,}\label{omega_B} \\ \noindent 
\omega_0^F(m, \mu) &=& -\frac{1}{2\pi^2\beta}\int^\infty_0dk\, k^2 \ln\left(1+{\rm
e}^{-\beta\left(\sqrt{k^2+m^2} - \mu \right)}\right) \text{.} \label{omega_F}
\end{eqnarray}  
\par For later convenience, the thermodynamic quantities will be normalized to the Stefan-Boltzmann pressure, which is defined as  
\begin{equation}
p_{SB} = - \displaystyle\lim_{m \rightarrow 0} \Omega_0^{\text{QCD}}\text{,}
\label{SB}
\end{equation}
\noindent and reads in this case
\begin{equation}\label{psbg}
p_{SB}=\frac{\pi^2}{45 \beta^4}  \left[ {\rm dim}\, adj \,  + \displaystyle\frac{7}{4}  \displaystyle\sum_{n = 1}^{N_f} {\rm dim}\, q_n \right]\text{.}
\end{equation}

\par As already mentioned, the second term of \eqref{pot0} stands for the interactions. The sum $\sum_\nu$ now explicitly reads $\sum_{n_g + n_{q_n} + n_{\bar{q}_n} = 2}\sum_{{\cal C}}\sum_{J^{P}}$, where $n_g$, $n_{q_n}$, $n_{\bar{q}_n}$ are respectively the number of gluons, quarks and antiquarks involved in the interaction process. Attractive interactions can lead to the formation of bound states with masses $M_{{\cal C},J^{P}}^{BS} < m_1+m_2$ (see Sec.~\ref{BS_QGP}). They contribute also to the grand potential as new species via the formula 
\begin{equation}
 \Omega_{bs}^{\text{QCD}} = \sum_{n_g + n_{q_n} + n_{\bar{q}_n} = 2} e^{\beta(\mu_1 + \mu_2)}
\sum_{J^{P}} (2J+1) \sum_{{\cal C}} {\rm dim}\, {\cal C}\,  \omega_0^{B/F}(M_{{\cal C},J^{P}}^{BS}, 0) \text{.}
\end{equation}
All $J^{P}$ and color channels leading to bound states are included in this summation. 
\par Concerning the scattering term, a tedious calculation (explained in \cite{LSCB, LSB}), leads to the following result,
\begin{eqnarray}
\Omega_s^{\text{QCD}} &=& \displaystyle\frac{1}{64 \pi^ 5\beta^ 2} \sum_{n_g + n_{q_n} + n_{\bar{q}_n} = 2}  e^{\beta(\mu_1 + \mu_2)} \displaystyle\sum_{J^P} (2 J + 1) \displaystyle\sum_{{\cal C}} \text{dim} {\cal C}  \\
&& \left(  \beta \displaystyle\int_{m_1 + m_2}^\infty d\epsilon \, \epsilon^2 \, \omega(\epsilon) \, \Lambda(\epsilon) K_1(\beta\epsilon) \,{\rm Re} T_{{\cal C}, J^P}(\epsilon; \omega(\epsilon), \omega(\epsilon)) \right.\nonumber \\ \nonumber
&& -\left. \displaystyle\frac{1}{16 \pi^2} \displaystyle\int_{m_1 + m_2}^\infty d\epsilon \, \epsilon^2 \, \omega(\epsilon)^2 \, \Lambda(\epsilon)^2 K_2(\beta\epsilon) \left[{\rm Re} T_{{\cal C}, J^P}(\epsilon; \omega(\epsilon), \omega(\epsilon)) \left({\rm Im} T_{{\cal C}, J^P}(\epsilon; \omega(\epsilon), \omega(\epsilon)) \right)' \right]  \right. \\ 
&& + \left. \displaystyle\frac{1}{16 \pi^2} \displaystyle\int_{m_1 + m_2}^\infty d\epsilon \, \epsilon^2 \, \omega(\epsilon)^2 \Lambda(\epsilon)^2  K_2(\beta\epsilon) \left[\left({\rm Re} T_{{\cal C}, J^P}(\epsilon; \omega(\epsilon), \omega(\epsilon))\right)' {\rm Im} T_{{\cal C}, J^P}(\epsilon; \omega(\epsilon), \omega(\epsilon))  \right]  \right)\text{,} \nonumber 
\end{eqnarray} 

\noindent where $\omega(\epsilon)$ and $\Lambda(\epsilon)$ are given by 
\begin{eqnarray}
\omega(\epsilon) &=& \displaystyle\frac{\displaystyle\sqrt{(\epsilon^2 - (m_1 + m_2)^2)(\epsilon^2 - (m_1 - m_2)^2)}}{2 \epsilon} \text{,} \\
\Lambda(\epsilon) &=& \displaystyle\frac{\epsilon^4 - (m_1^2 - m_2^2)}{\epsilon^3} \text{,} \label{lambda_eps}
\end{eqnarray}
and where $T_{{\cal C}, J^P}(\epsilon; \omega(\epsilon), \omega(\epsilon)) $ is the on-shell $T_{{\cal C}, J^P}$-matrix. Note that a isospin number $I$ has to be taken into account when one deals with $u$ and $d$ quarks since they have the same mass in our approach. This isospin number enters in the summation $\sum_{n_g + n_q + n_{\bar{q}}=2}$ as a $(2I + 1)$ factor.

\par Finally, the grand canonical potential (reduced to two-body interactions) is summarized by the following formula

\begin{equation} \label{gen}
\Omega_{(2)}^{\text{QCD}} = \Omega_{0}^{\text{QCD}} + \Omega_{bs}^{\text{QCD}} + \Omega_{s}^{\text{QCD}} \text{.}
\end{equation}

\noindent For obvious numerical reasons, the summation over the number of particles is not the only one that must be restricted. All possible color channels are included, but all the possible $J^{P(C)}$ channels contributing to $\Omega_{(2)}^{\text{QCD}}$ can not be included since their number is infinite.  So, a reliable criteria to select the most significant ones has to be established. The basic idea, already proposed in \cite{LSB}, is that only states with low $L$ are included since they are the most likely to contribute significantly to a total mean cross section $\bar \sigma_{J^P}$.  Are only retained, the channels for which the value of $\bar \sigma_{J^P}$ is at least 25\% of the value $\bar \sigma_{J^P}$ for the channel with the lowest value of $\left\langle \vec L^2\right\rangle$ (see Appendix~D of \cite{LSB} for further explanations).
\par In the present case, this criterion implies that only the following $J^{P(C)}$ channels are included.
\begin{itemize}
\item For $gg$ channels: the $0^{++}$, $0^{-+}$, $2^{++}$ and $1^{++}$ states;
\item For $qq$, $\bar{q}\bar{q}$ and $q\bar{q}$ channels: all the $J^P$ ones with $L = 0$ or $L = 1$;
\item For the $qg$ and $\bar{q}g$ channels: all the states with $\langle \vec{L}^{2} \rangle < 8$ (see Appendix~B in \cite{LSB}). 

\end{itemize}

\subsection{QGP with $\bm{N_f = 2}$}

\par
\begin{figure}[h!]
\begin{center}
\includegraphics*[width=0.49\textwidth]{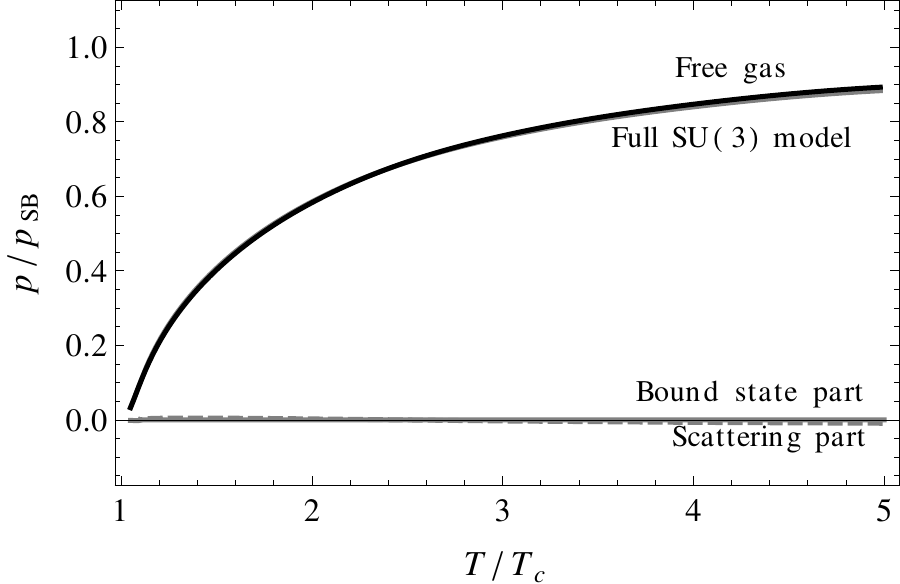}
\includegraphics*[width=0.49\textwidth]{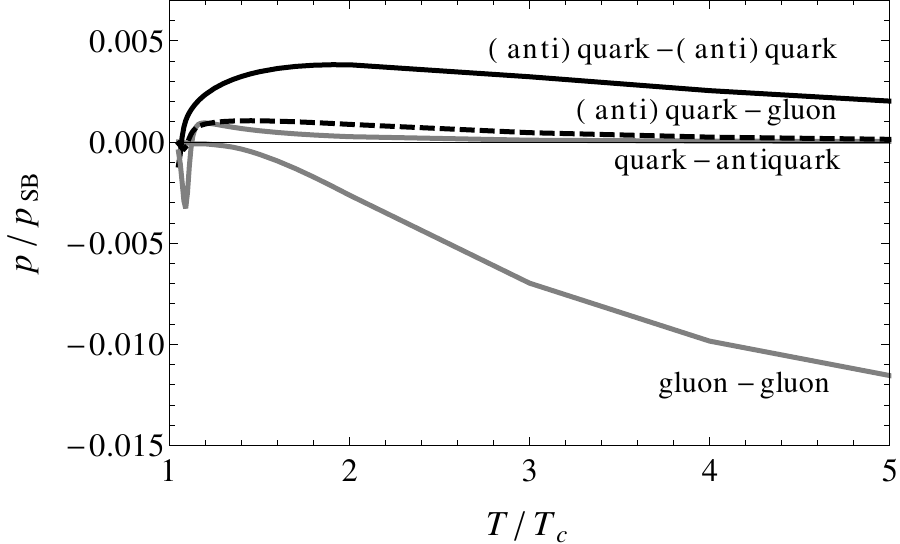}
\caption{(Left) Normalized pressure $p/p_{SB}$ versus temperature in units of $T_c$, compared to the free part, bound state and scattering contribution. (Right) Different scattering contributions to the normalized pressure $p/p_{SB}$ versus temperature in units of $T_c$. $T_c = 0.15$~GeV.}
\label{QGP2_pressure}
\end{center}
\end{figure}

 Now that the number of $J^P$ channels for each two-body interactions is fixed, the EoS can be computed. In Fig.~\ref{QGP2_pressure}, the normalized pressure is shown for a QGP with two light-quark flavor included. As it can be noticed, interactions do not practically contribute: The major part is given by the free gas. Globally, the weakness of interactions can be interpreted in the same way as what it is observed for the YM plasma in Sec.~\ref{param_Tc}. Indeed, when the critical temperature decreases, the interactions seem to become smaller and smaller. This behaviour is driven by the Bessel functions entering in the definition of the scattering part. Moreover, it is important to notice that the integration range in this term formally starts at $\beta (m_1 + m_2)$. This value is large in comparison to the values at which the Bessel functions is significantly non-zero. 
A change of the thermal mass prescription could thus impact the contributions of the scattering part. This work is left for further developments of our approach. 
\par In the right panel of Fig.~\ref{QGP2_pressure}, the different scattering contributions are separated. Without surprise, the $q\bar{q}$, $qg$ and $\bar{q}g$ channels asymptotically tend to zero. Indeed, it has been shown in \cite{LSCB} that the interactions between two different species vanish within the Born approximation, because of an identity relating the color factors: $\sum_{{\cal C}}{\rm dim}\,{\cal C} \ \kappa_{{\cal C},ij}=0$. Concerning the $qq$ and $\bar{q}\bar{q}$ channels, they generate a global increase of the normalized pressure while it is the contrary for the $gg$ sector. Not only these two effects are weak but in addition, they contribute in opposite directions, leading to a global suppression of the two-body interactions in average.

\begin{figure}[h!]
\begin{center}
\includegraphics*[width=0.49\textwidth]{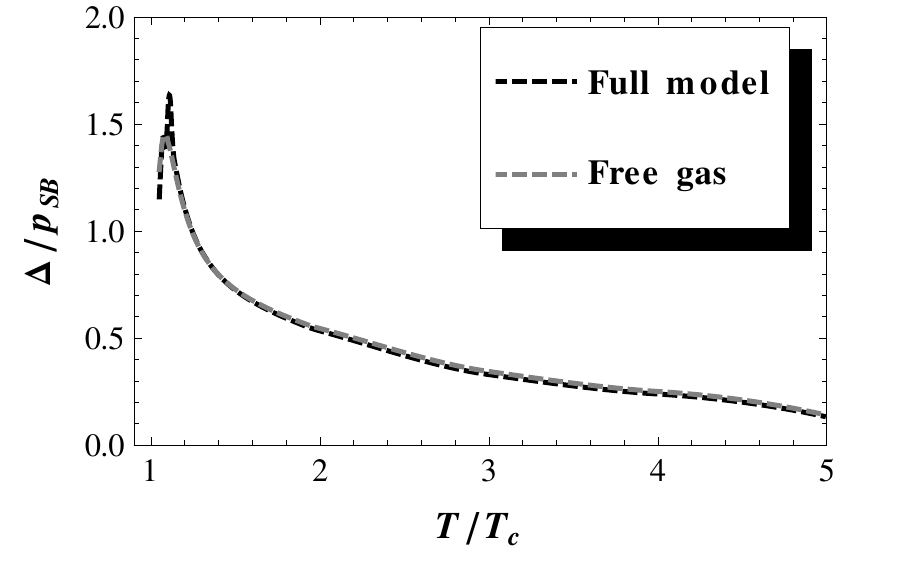}
\caption{Normalized trace anomaly $\Delta/p_{SB}$ (without bound states) versus temperature in units of $T_c$, compared to the free part contribution. $T_c = 0.15$~GeV.}
\label{QGP2_anom}
\end{center}
\end{figure}

\par In Fig.~\ref{QGP2_anom}, we display the normalized trace anomaly (without bound states) compared to the free gas part. A peak structure is here exhibited even in the free gas contribution. Therefore, it is different from the YM sector where the interactions create the peak. The nature of this latter is really difficult to establish since few variations of the pressure can drastically change the shape of the trace anomaly. 
\par The main conclusion that seems to emerge from our approach (looking at the normalized trace anomaly as well as at the normalized pressure) is that the leading behavior of the QGP is driven by gluon and (anti)quark degrees of freedom that interact weakly. Nevertheless, it does not mean that the interactions have no impact on the EoS. Indeed, the particle thermal mass is extracted from it, leading to a self-energy contribution for the particle (see Sec.~\ref{QGP_3}).

\subsection{QGP with $\bm{N_f = 2 + 1}$} \label{QGP_3}

\par A similar analyse as the one proposed in the previous subsection can be applied in the 2 + 1 QGP case. Since similar results and features can be deduced from it, we will not repeat it again and focus more on the comparisons between our model and lQCD extracted from \cite{Bors14}. Indeed,  lQCD collaborations have recently reached the physical quark masses in their computations of the EoS, making their results more and more reliable for comparisons. 

\begin{figure}[h!]
\begin{center}
\includegraphics*[width=0.49\textwidth]{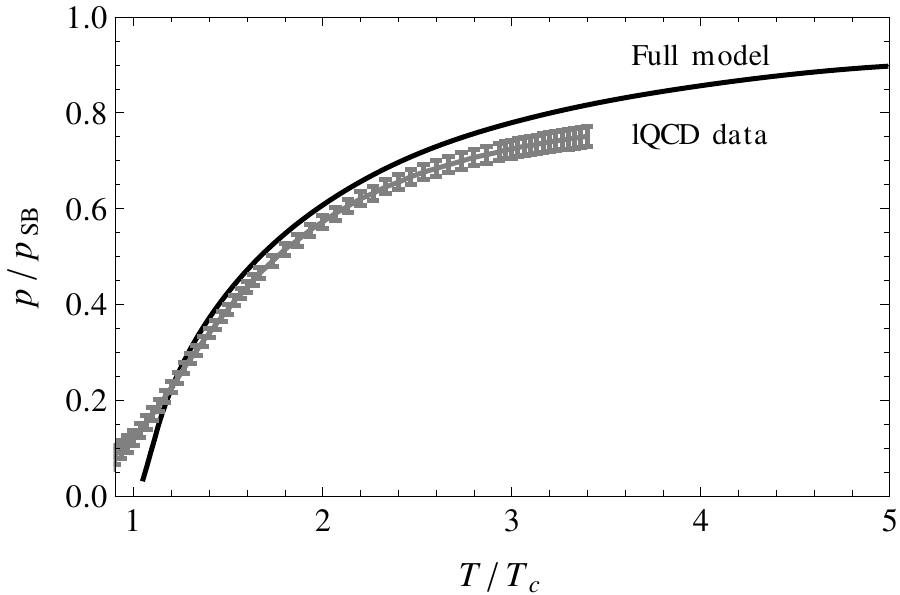}
\includegraphics*[width=0.49\textwidth]{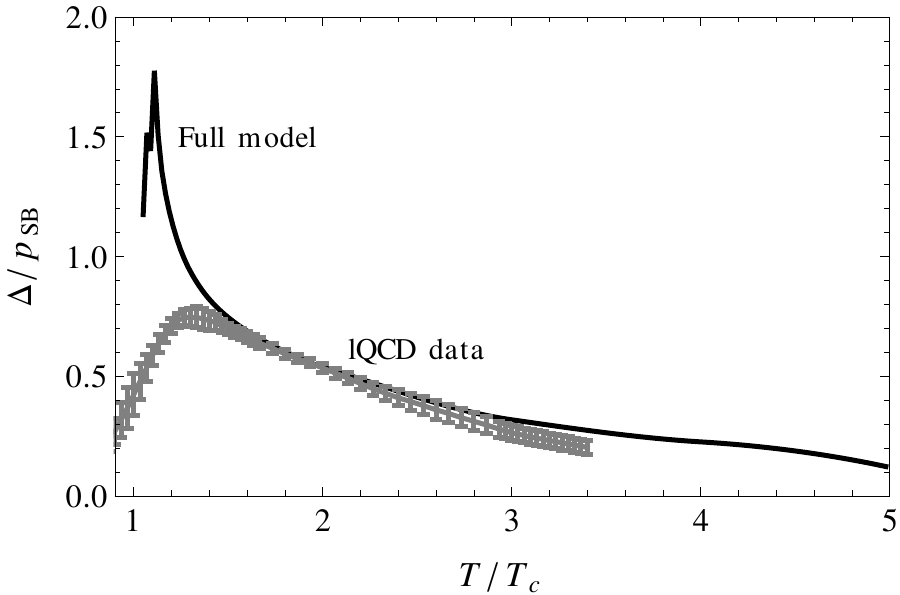}
\caption{(Left) Normalized pressure $p/p_{SB}$ versus temperature in units of $T_c$, compared to lQCD data from \cite{Bors14}. (Right)  Normalized trace anomaly $\Delta/p_{SB}$ versus temperature in units of $T_c$, compared to lQCD data from \cite{Bors14}. $T_c = 0.15$~GeV.}
\label{QGP21}
\end{center}
\end{figure}

\par As we can observe in Fig.~\ref{QGP21}, our data are qualitatively in agreement with lQCD ones. The lQCD normalized pressure is slightly overestimated as well as the asymptotic behavior of the normalized trace anomaly. On the other hand, the peak structure of the trace anomaly is very different of the lQCD one. As already mentioned, this latter is really difficult to obtain due to several reasons in our approach: problems in the inclusion of the bound state \cite{LSCB}, reliability of the quark masses and restriction to two-body interactions. Even in lQCD, different collaborations find different quantitative behaviors for the trace anomaly peak \cite{Bors14}. The possible discrepancies can arise from the choice of the fermionic lattice action, the lattice spacing, the considered quark masses, the extrapolation to the continuum limit... Only, a good agreement in the behavior of the decreasing tail is reached by the different lQCD groups, according to \cite{Bors14}. Nevertheless, it is worth mentioning that the disagreements  observed in the quantitative value of the peak structure in various lQCD results (mainly due to a computation with no physical quark masses) seem to reduce, and the shape of the lQCD trace anomaly tends to the one depicted in Fig.~\ref{QGP21}, and firstly given by the BMW collaboration \cite{Ratt14}. 

\par
\begin{figure}[h!]
\begin{center}
\includegraphics*[width=0.49\textwidth]{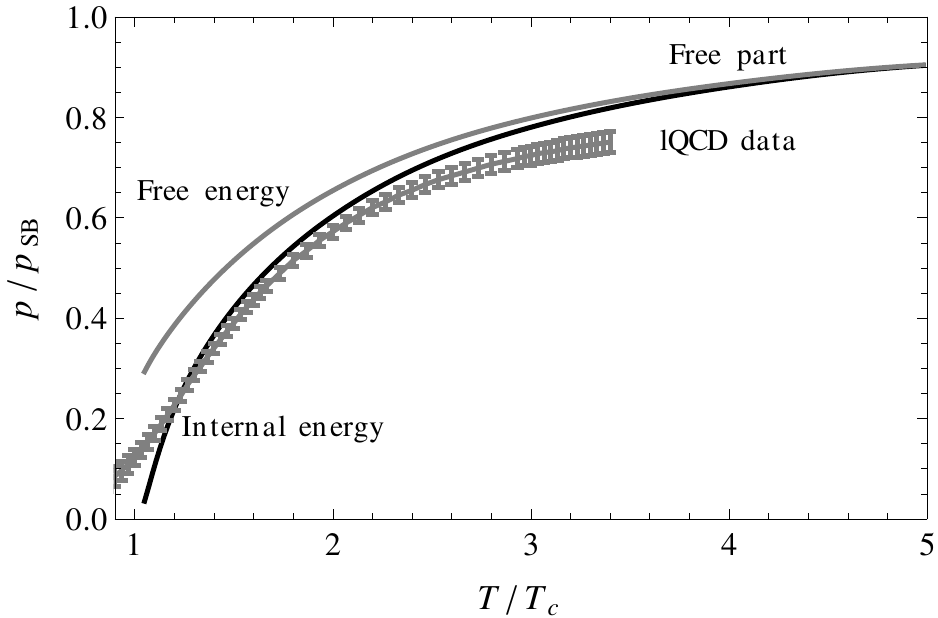}
\includegraphics*[width=0.49\textwidth]{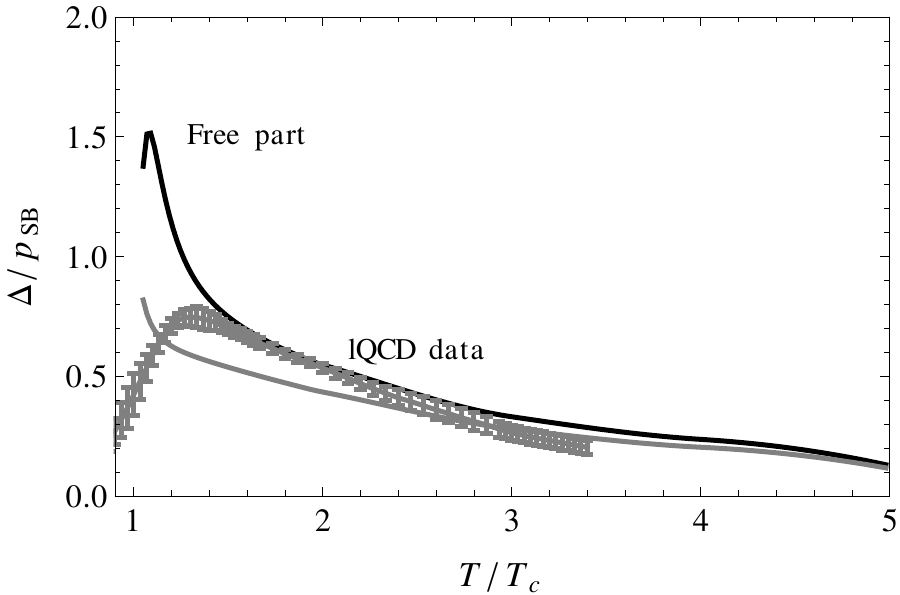}
\caption{(Left) Normalized pressure $p/p_{SB}$ versus temperature in units of $T_c$. (Right) Normalized trace anomaly $\Delta/p_{SB}$ (without bound states) versus temperature in units of $T_c$. In the two figures, the gray (black) line is the free part contribution of $\Omega_{(2)}^{\text{QCD}}$ computed with the free (internal) energy. $T_c = 0.15$~GeV.  }
\label{UvsF}
\end{center}
\end{figure}

 Therefore, except for the normalized trace anomaly peak structure (for which a more appropriate treatment of the bound-state inclusion is needed and could change significantly its structure), our data are in correct agreement with lQCD ones. As discussed in the previous subsection, this agreement seems to be reached by only including a quasiparticle thermal mass: The contributions of the two-body interacting channels are minor. Nevertheless, it is worth insisting on the fact that the thermal mass effects are extracted  from the two-body lQCD interaction potential within our model. So, the chosen two-body interactions are not useless to understand the behavior of the QGP around $T_c$. Indeed, if we change the potential, the free gas contribution is modified since the quasiparticle thermal masses depend on it. This leads to a completely different behavior of the EoS as seen in Fig.~\ref{UvsF} in which the potential is now chosen to be the free energy. 
We can especially notice in Fig.~\ref{UvsF} that a better agreement between our model and lQCD normalized pressure is reached around $T_c$ thanks to the internal energy while the discrepancy between the two curves decrease when the temperature increases. 
\par
\begin{figure}[h!]
\begin{center}
\includegraphics*[width=0.49\textwidth]{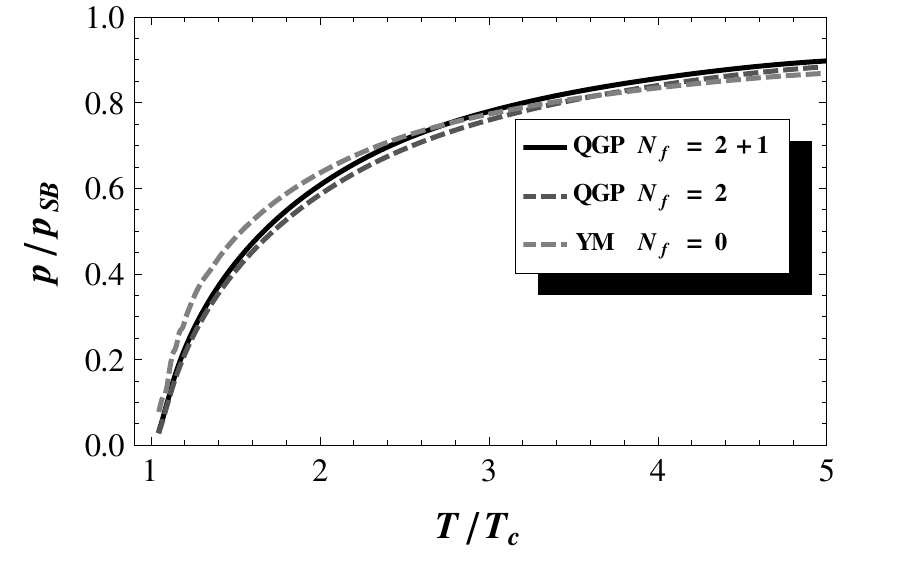}
\includegraphics*[width=0.49\textwidth]{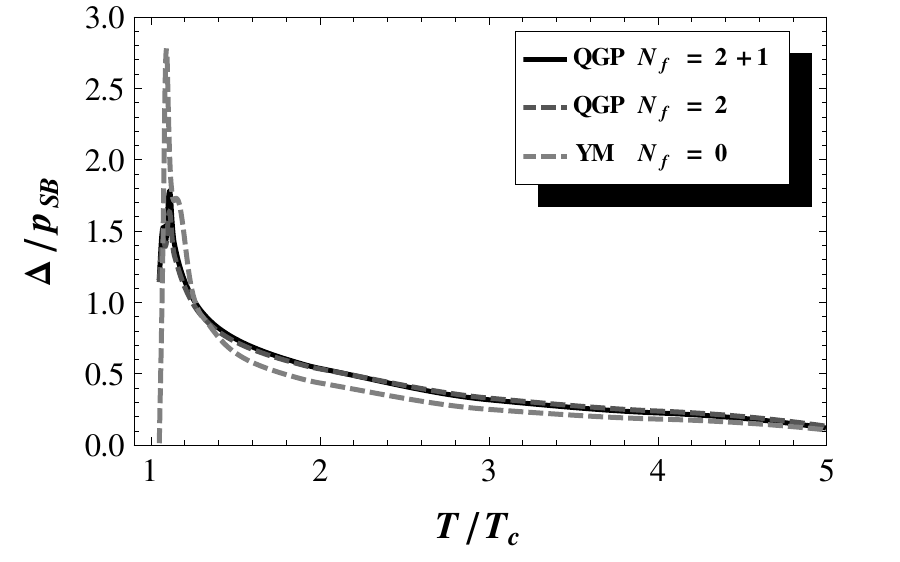}
\caption{(Left) Normalized pressure $p/p_{SB}$ versus temperature in units of $T_c$. (Right) Normalized trace anomaly $\Delta/p_{SB}$ (without bound states) versus temperature in units of $T_c$. $T_c = 0.15$~GeV for QGP and $T_c = 0.3$~GeV for YM. }
\label{QGP21vsQGP2}
\end{center}
\end{figure}

 Finally, let us compare in Fig~\ref{QGP21vsQGP2} the normalized pressure and trace anomaly for a QGP with $N_f = 2$ and $N_f = 2 + 1$ ($T_c =$ 0.15 GeV) to the ones of the YM plasma ($T_c =$ 0.3 GeV). We can notice that the normalized pressure curves are almost superimposed and that the decreasing trend of the trace anomaly is nearly the same in all the considered theories. The maximum of the deviation between these curves is around $1.2\, T_c$, at the localisation of the trace anomaly peak. It is nevertheless important to remember that the critical temperature and the normalization are not the same in all the EoS (see \eqref{psbg}). However, within these units, a universality at large temperature ($\geq 3\, T_c$) seems to emerge. 

\section{Equation of state of the QGP at small $\mu$} \label{EoS_QGP_mu}

\par Now that the EoS for the QGP are computed and favourably compared with lQCD, we can investigate the non-zero baryonic regime. This latter deserves a lot of interests, especially in the area of the neutron star physics. Indeed, since pioneering works \cite{Coll75} about the existence of a deconfined phase in QCD, it was assumed that the core of the heaviest neutron star should be probably filled by a medium with a high nuclear density and in which the significant degrees of freedom should be the quarks. Therefore, getting the QCD EoS at finite $\mu$ could shed some light in this field. 
\par Up to now, this task still remain difficult from first QCD principles. Remember that even in lQCD some conceptual troubles appear (cfr.~sign problem) and only perturbations around $\mu = 0$ are meaningful. Therefore, it seems appealing to check whether or not quasiparticle approaches could help. Unfortunately at the present stage, some problems also appear in our formalism.  The main reasons are the following. 
\par First, the Dashen Ma and Bernstein formalism that we have used to compute the EoS is based on a virial expansion in terms of $e^{\beta \vec{\mu} \vec{N}}$. We are thus limited by construction to small baryonic potential. Indeed, increasing the baryonic potential is the same as increasing the density of particles: The many-body interactions are more and more likely to contribute. So, the reduction to two-body interactions becomes a poor approximation \textit{a priori} and some problems, other than a careful computation of all the channels, arise. Let us mention for instance, the absence of a helicity formalism for many-body systems in a potential approach, and the necessity to resort to Faddeev and higher equations for more than two-body interactions. Moreover, when the density of particles increases, the notion of quasiparticle becomes more and more questionable.
\par Another peculiar problem is the building of a coherent interaction in presence of baryonic potential. Already at two bodies, no lQCD data are available to our knowledge. It is not only important to define the potential between particles but also the quasiparticle mass, which seems to rule the main behavior of the EoS at $\mu = 0$. A way to circumvent this problem could be to use the HTL expressions for the particle thermal mass, but it was not the bias adopted within this study. Indeed, the actual shape of our thermal masses are not the ones extracted from HTL. 
\par
\begin{figure}[h!]
\begin{center}
\includegraphics*[width=0.32\textwidth]{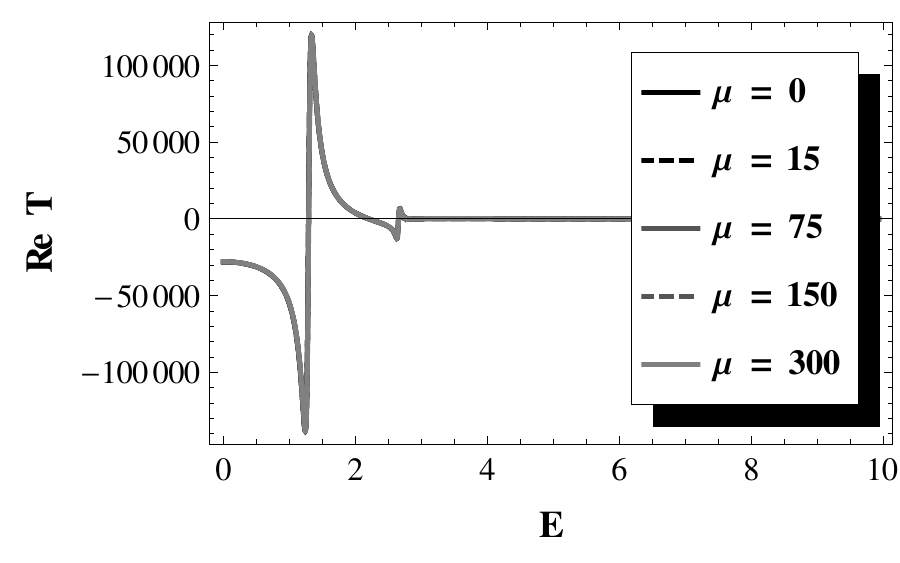}
\includegraphics*[width=0.32\textwidth]{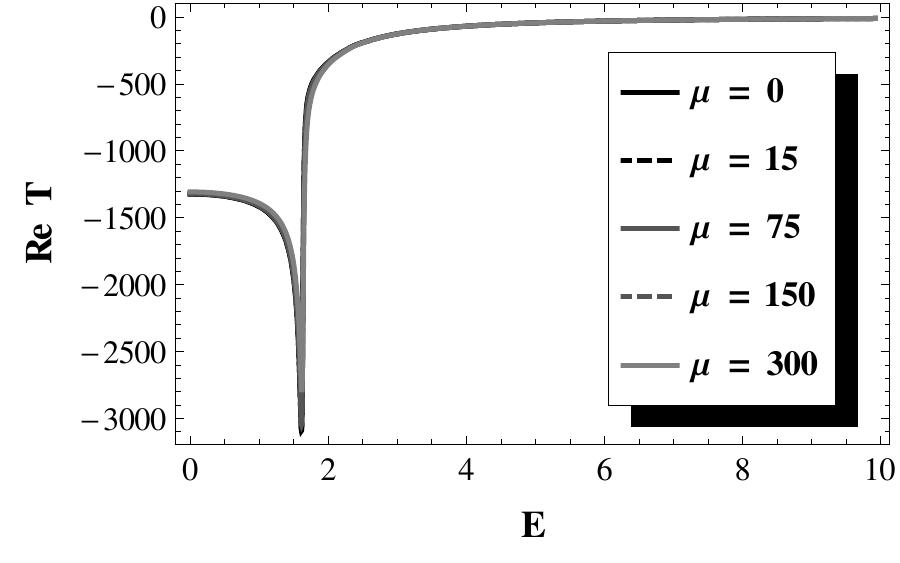}
\includegraphics*[width=0.32\textwidth]{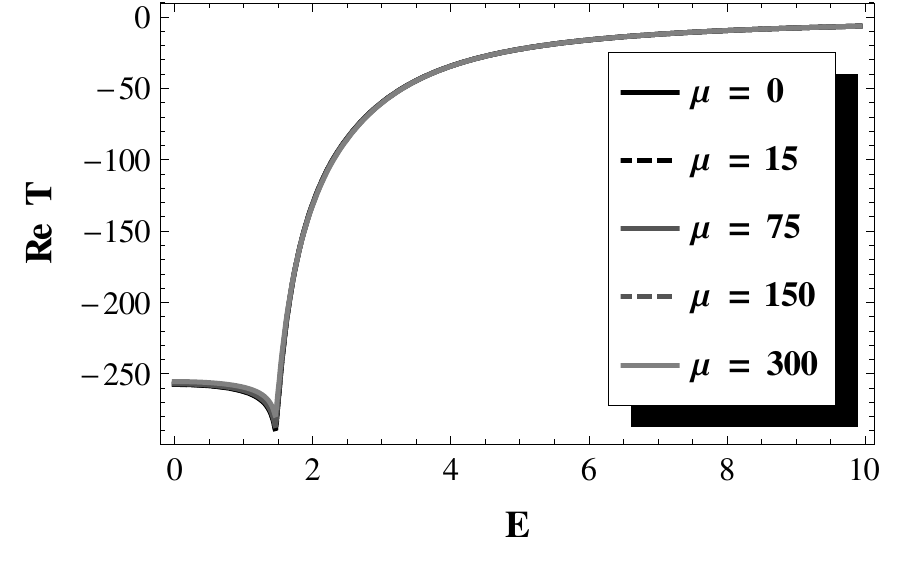}
\caption{Real part of the $T$-matrix ($q\bar{q}$ in $L = 0$, here given as example) in function of the energy for $T$ = 0.105 GeV (left), $T$ = 0.150 GeV (middle) and $T$ = 0.300 GeV (right) at different $\mu$ (MeV) with $T_c$ = 0.15 GeV. }
\label{mu_impact}
\end{center}
\end{figure}

 For all these reasons, the study that follows will be only limited to small baryonic potentials. We will thus keep the restriction to two-body interactions which can make sense in such a $\mu$-range. Moreover, the interaction potential and the quasiparticle thermal masses are the same as the ones used up to now, without the inclusion of the baryonic potential. Of course, the obtained results must be considered as preliminary and are just intended to draw a general tendency. The baryonic potential enters at two levels in our computations: in the $T$-matrices because of the in-medium effects, and in all the EoS contributions as multiplicative factors. Fortunately as for the $T_c$-impact, it seems that the $\mu$-dependence on the $T$-matrix calculations is negligible (see Fig.~\ref{mu_impact}). Therefore, these latter do not have to be recomputed at each $\mu$, which drastically reduces the computational time.

\par
\begin{figure}[h!]
\begin{center}
\includegraphics*[width=0.49\textwidth]{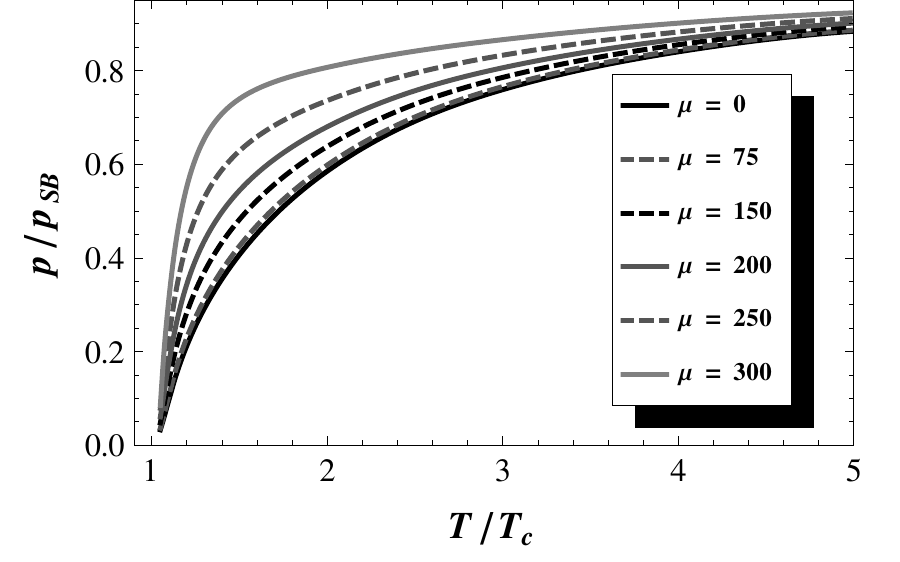}
\includegraphics*[width=0.49\textwidth]{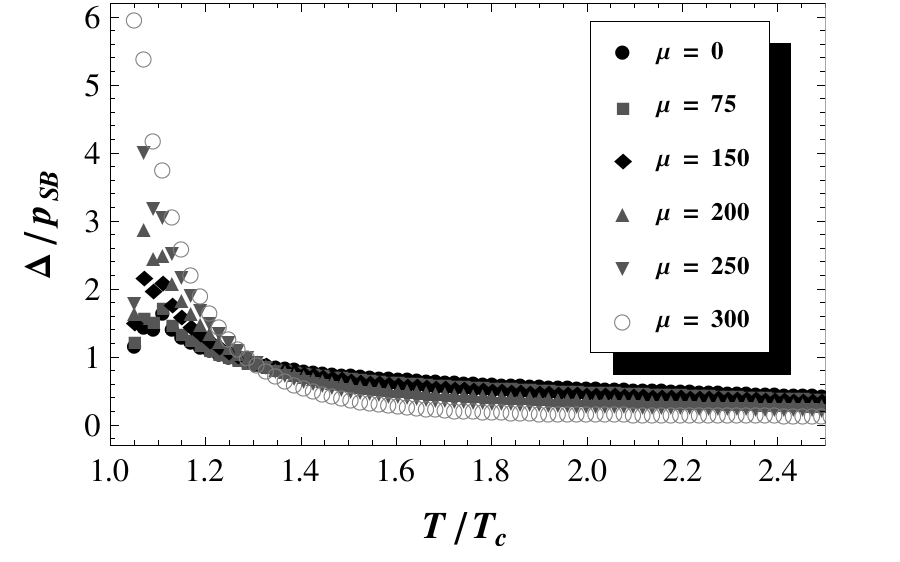}
\caption{(Left) Normalized pressure $p/p_{SB}$ versus temperature in units of $T_c$ at different $\mu$ (MeV) for a QGP with two light quarks. (Right) Normalized trace anomaly $\Delta/p_{SB}$ (without bound states) versus temperature in units of $T_c$  at different $\mu$ (MeV) for a QGP with two light quarks. $T_c = 0.15$~GeV.}
\label{pressure_mu}
\end{center}
\end{figure}

\begin{figure}[h!]
\begin{center}
\includegraphics*[width=0.49\textwidth]{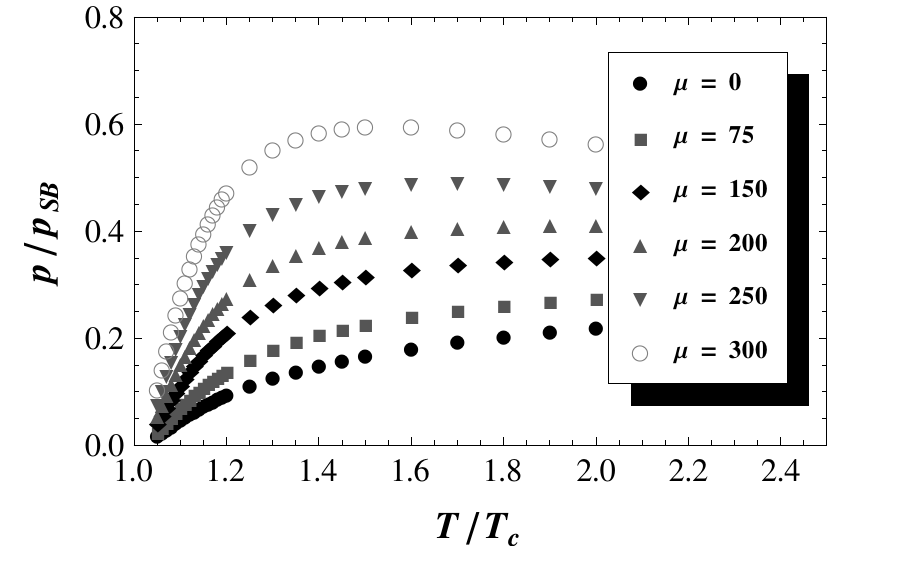}
\includegraphics*[width=0.49\textwidth]{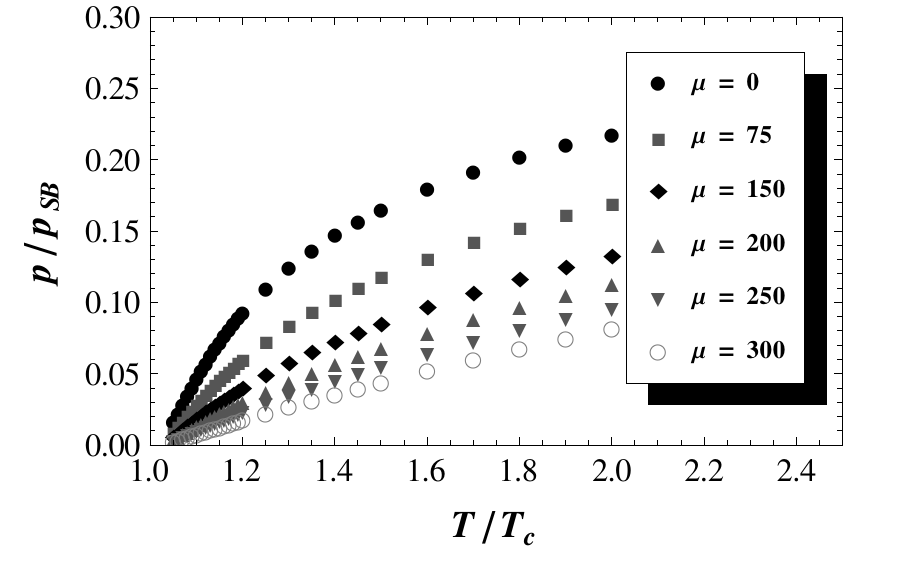}
\caption{(Left) Free quark gaz contribution to the total normalized pressure $p/p_{SB}$ versus temperature in units of $T_c$ at different $\mu$ (MeV) for a QGP with two light quarks. (Right) Free antiquark gaz contribution to the total normalized pressure $p/p_{SB}$ versus temperature in units of $T_c$ at different $\mu$ (MeV) for a QGP with two light quarks. $T_c = 0.15$~GeV.}
\label{interaction_mu1}
\end{center}
\end{figure}

\begin{figure}[h!]
\begin{center}
\includegraphics*[width=0.40\textwidth]{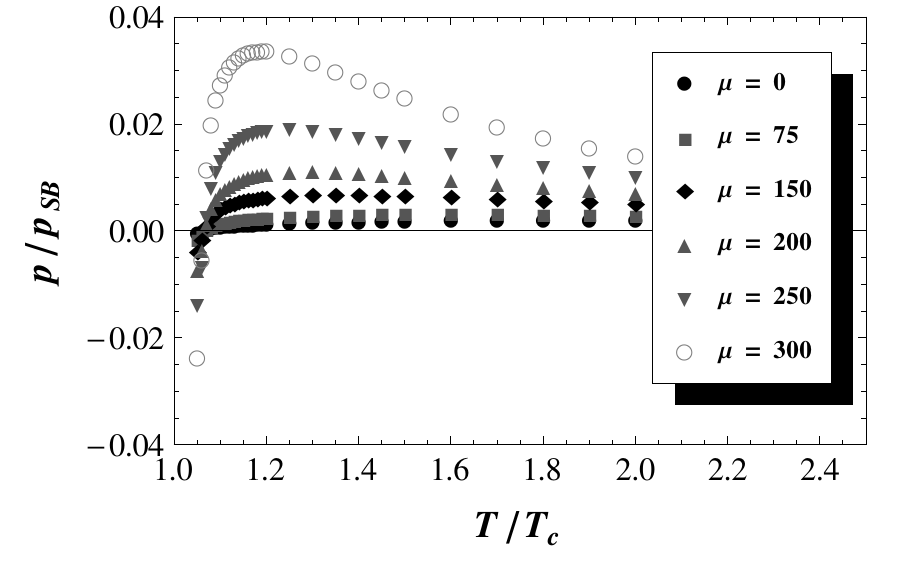}
\includegraphics*[width=0.40\textwidth]{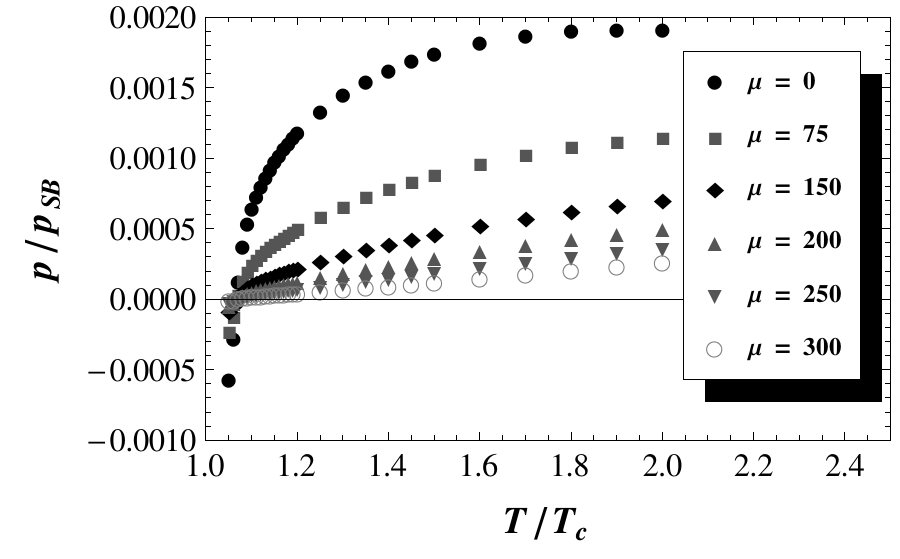}
\includegraphics*[width=0.40\textwidth]{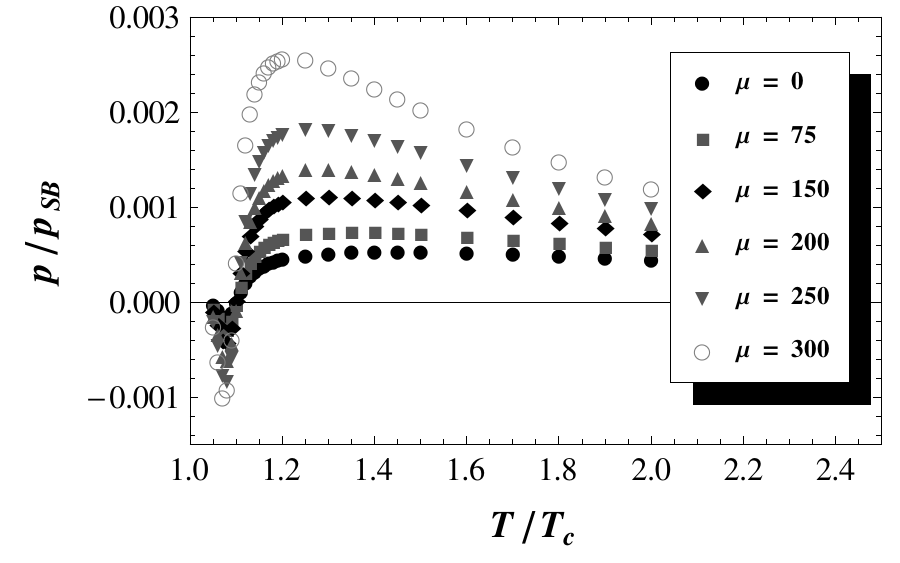}
\includegraphics*[width=0.40\textwidth]{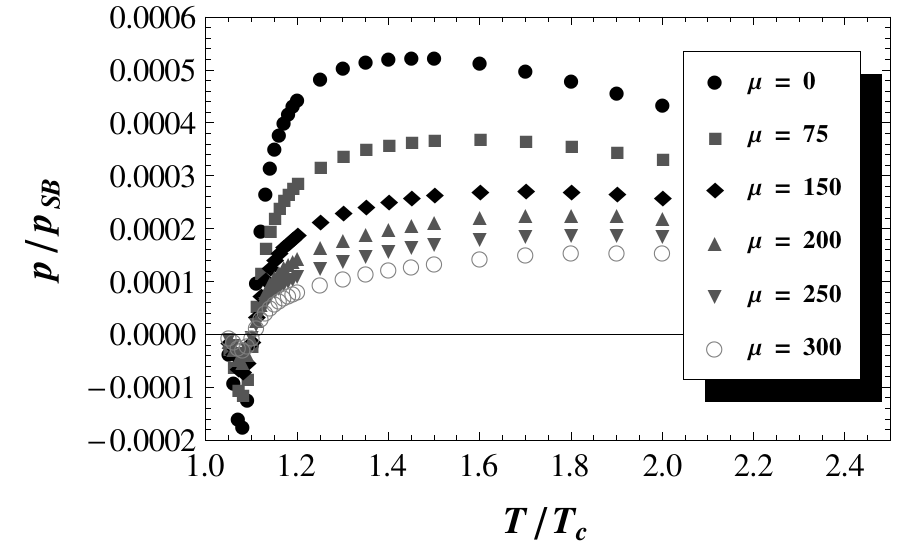}
\includegraphics*[width=0.40\textwidth]{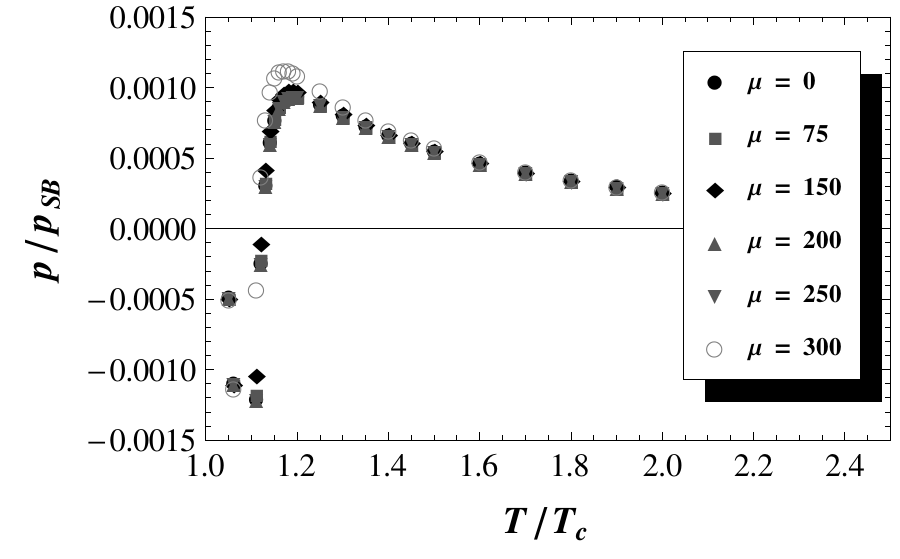}
\caption{(Left, top) $qq$-scattering contribution to the total normalized pressure $p/p_{SB}$. (Right, top) Same for $\bar{q}\bar{q}$. (Left, middle) Same for $qg$. (Right, middle) Same for $\bar{q}g$. (Bottom) Same for $q \bar{q}$. All the scattering contributions are presented versus temperature in units of $T_c$, with $T_c = 0.15$~GeV, at different $\mu$ (MeV) for a QGP with two light quarks.}
\label{interaction_mu2}
\end{center}
\end{figure}

 In Fig.~\ref{pressure_mu}, we have plotted the normalized pressure and trace anomaly at different $\mu$ for a QGP with two light quarks. The normalization is given by \eqref{psbg}, that is to say at $m = 0$ and $\mu = 0$. Naturally, the gluon chemical potential is zero and the quark one is such that $\mu_u = \mu_d = \mu$. We can see in this figure that the normalized pressure increases with $\mu$. This pressure is especially driven by the increase of the free quark gas contribution given in Fig.~\ref{interaction_mu1}. Indeed, as in the $\mu = 0$ case, the leading contributions to the normalized pressure are the free part ones since the impact of the interactions is small as observed in Fig.~\ref{interaction_mu2}. Moreover, the decrease of the free antiquark gas contribution is slower than the increase of the free quark gas one, explaining the total increasing behavior of the normalized pressure.

\par Concerning the normalized trace anomaly, it is much more difficult to understand the $\mu$-dependence. The only assertion that we can do is that the trace anomaly peak becomes higher and higher with the increase of $\mu$. Moreover, we can notice that the convergence to zero is faster with large $\mu$. 

\par As already mentioned, the scattering contributions are small. Nevertheless in Fig.~\ref{interaction_mu2}, we can observe a significant dependence in terms of  $\mu$. The $qq$ and $qg$ scattering contributions obviously increase with $\mu$, respectively as $e^{2 \beta \mu}$ and $e^{\beta \mu}$, while the $\bar{q}\bar{q}$ and $\bar{q}g$ ones go in opposite way. However, as in the free gas case, the increase is higher than the decrease, leading \textit{in fine} to a more important contribution of the scattering parts to the total normalized pressure. To be complete, the $q\bar{q}$ scattering contribution is stable since there is no $\mu$-dependence at the level of the EoS: Indeed, we have $e^{\beta(\mu - \mu)} = 1$ and just a very weak dependence on $\mu$ appears in the $T$-matrix. Moreover, since the $gg$ sector is independent of $\mu$, increasing $\mu$ means increasing the impact of the quark sector within the QGP.

\par Finally, we close this study by comparing our preliminary results to the lQCD ones given by \cite{Bors12b}. Within this paper, they deal with a QGP with $N_f = 2 + 1$ and with a small baryonic potential $\mu_B$. Each flavor of quarks is considered to carry one third of $\mu_B$. Therefore, we analyse the 2 + 1 QGP with $\mu_u = \mu_d = \mu_s = \mu_B/3$. As for the $\mu = 0$ case, we sightly overestimate the normalized pressure and we miss the peak of the normalized trace anomaly. Therefore, it seems that these differences have mainly the same origin as at $\mu = 0$ and our extrapolations at small $\mu$ is compatible with lQCD. 

\begin{figure}[h!]
\begin{center}
\includegraphics*[width=0.40\textwidth]{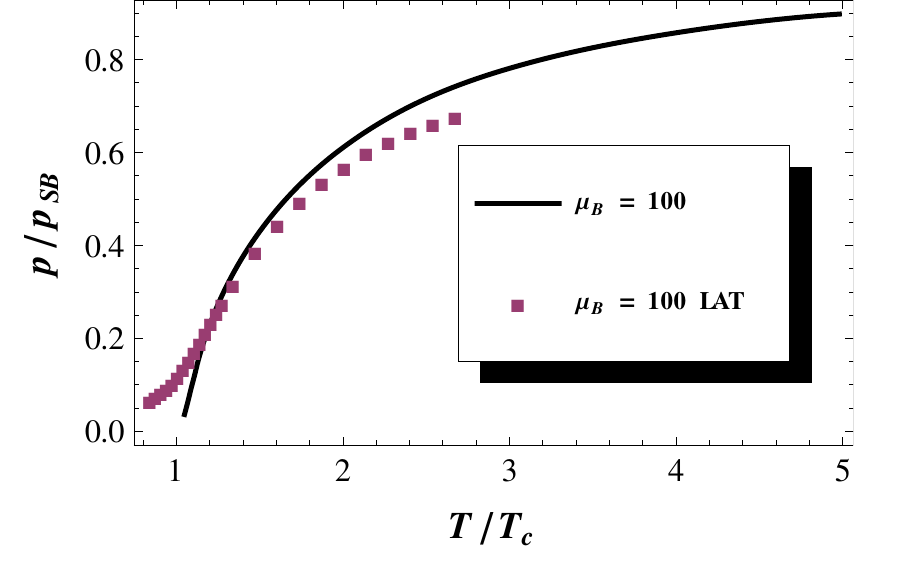}
\includegraphics*[width=0.40\textwidth]{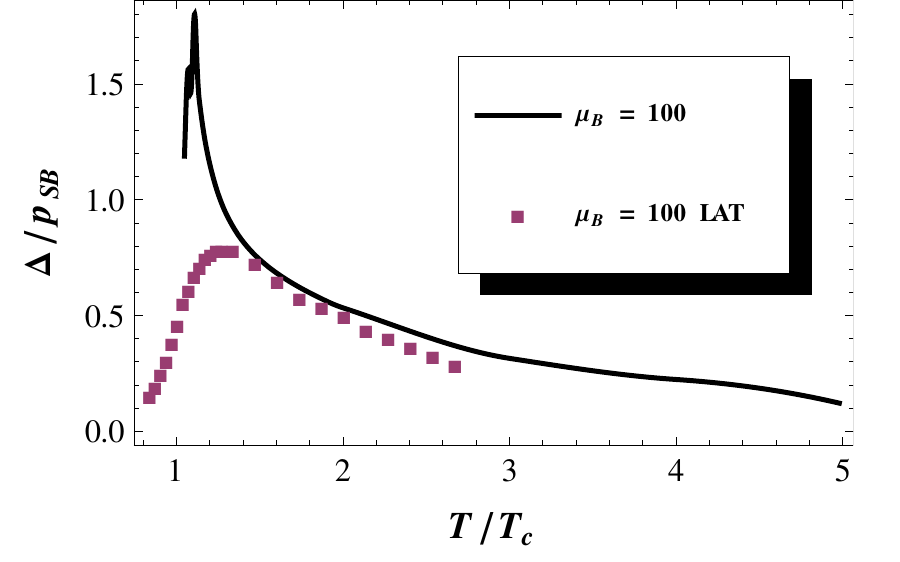}
\includegraphics*[width=0.40\textwidth]{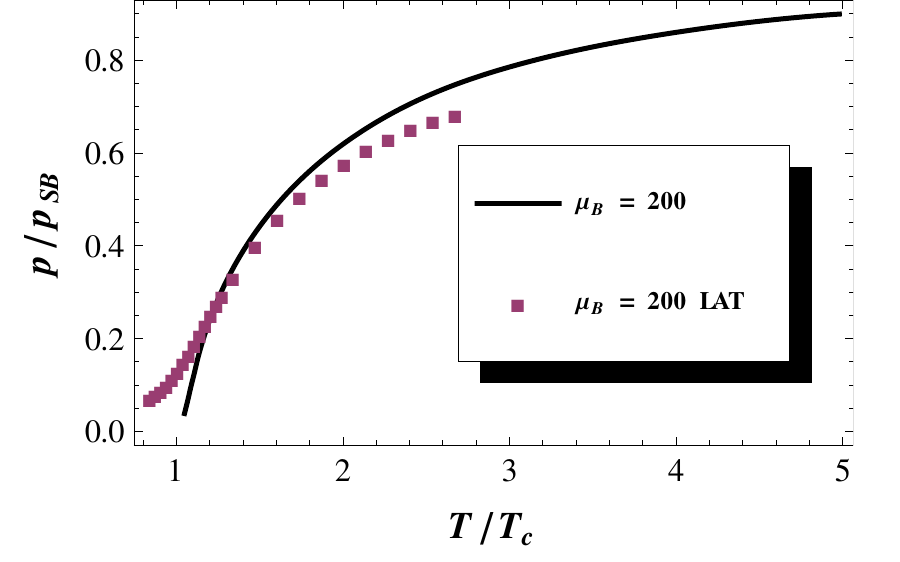}
\includegraphics*[width=0.40\textwidth]{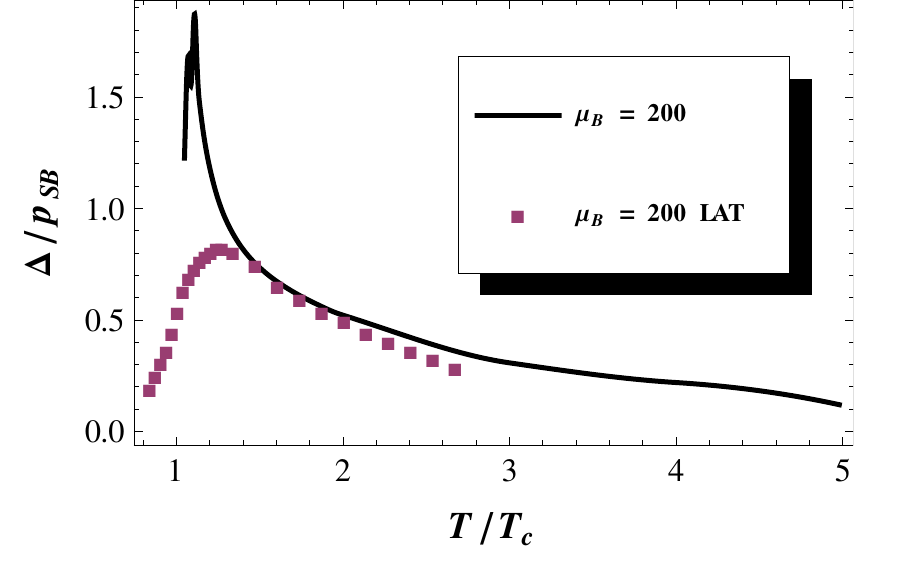}
\includegraphics*[width=0.40\textwidth]{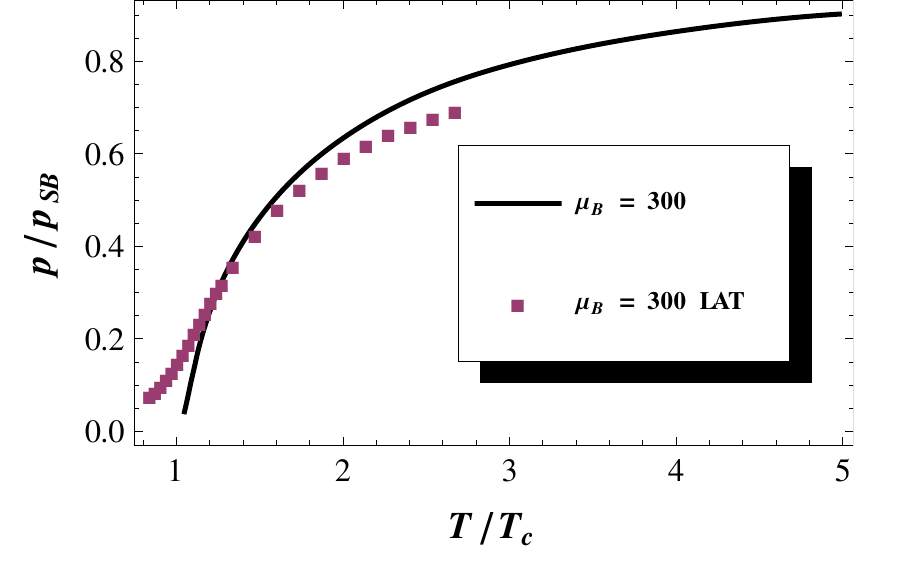}
\includegraphics*[width=0.40\textwidth]{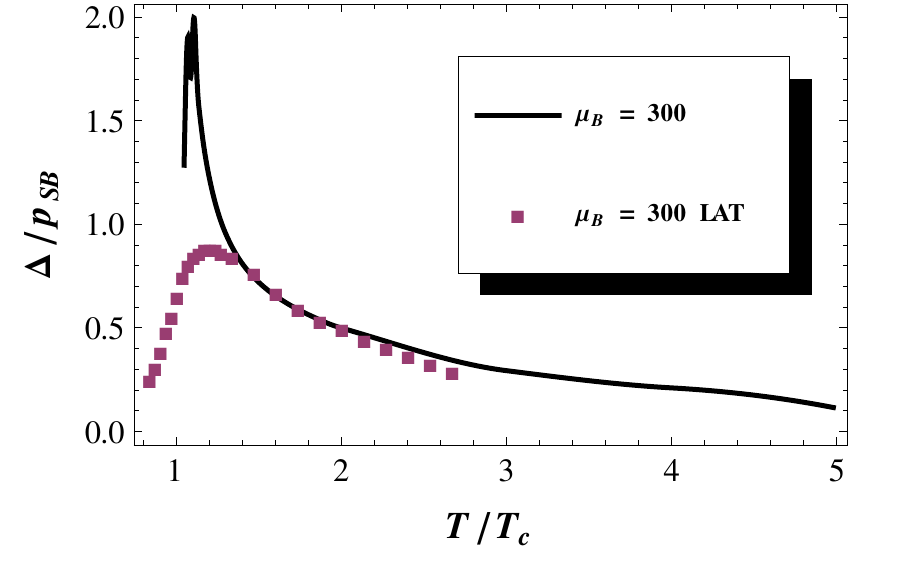}
\includegraphics*[width=0.40\textwidth]{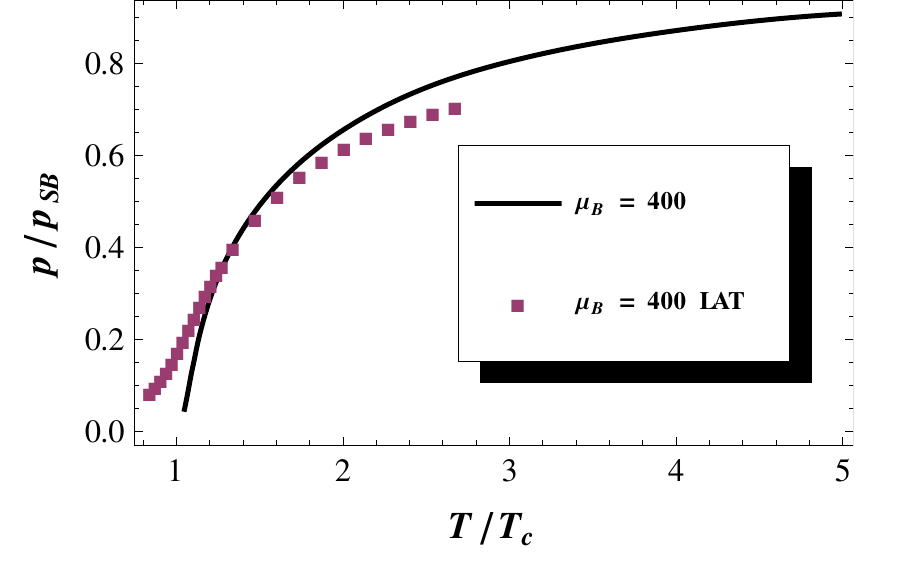}
\includegraphics*[width=0.40\textwidth]{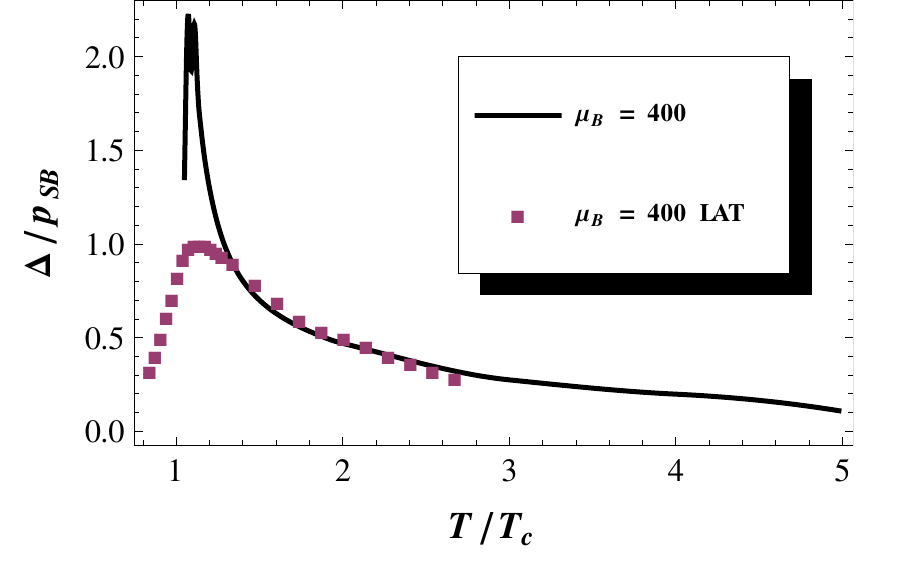}

\caption{(Color online) (Left) Normalized pressure $p/p_{SB}$ compared to lattice QCD \cite{Bors12b} versus temperature in units of $T_c$ at different $\mu_B$ (MeV) for a QGP with $N_f = 2 + 1$. (Right) Normalized trace anomaly $\Delta/p_{SB}$ compared to lattice QCD \cite{Bors12b} versus temperature in units of $T_c$ at different $\mu_B$ (MeV) for a QGP with $N_f = 2 + 1$. $T_c = 0.15$~GeV.}
\label{interaction_mu3}
\end{center}
\end{figure}

\section{Conclusions}\label{conclu}
The present work is part of a program aiming at studying the thermodynamic properties of gauge theories in the deconfined phase. The interested reader may read \cite{LSCB} and \cite{LSB} for pure YM and SUSY YM theories, while this paper is devoted to the ``realistic" quark-gluon plasma. The framework developed is based on a $T$-matrix formulation of statistical mechanics, in which the thermal masses and two-body interactions are derived from the static potential between fundamental color source computed in quenched lattice QCD \cite{lPot}. Apart from the potential, the only remaining parameters are the value of $T_c$ and the bare quark masses. These masses are fitted on the meson spectrum at zero temperature. The main assumption underlying our model is actually that a quasiparticle picture of deconfined matter just above deconfinement is relevant. Although it is not a rigorous proof, the nice agreement between our computed equations of state an the recent lattice data of Refs. \cite{Bors12b,Bors14} can be seen as an \textit{a posteriori} validation of our framework.

We are now in position of summarizing some of the key results obtained in this paper:
\begin{itemize}
\item Both the free energy or the internal energy could be used as potential terms in our model. It appears that, keeping the same procedure and the same values for the parameters, only the internal energy is able to generate an equation of state which has the qualitative features of the lattice equation of state. The internal energy thus appears as the most relevant potential within in our framework and leads to a good agreement with the lattice equation of state. Note that this problem is far from being elucidated, see for example the recent work \cite{Satz}, where the opposite conclusion is reached.
\item Between 1 and 2 $T_c$, color interactions are strong enough to create mesons, \textit{i.e.} a quark-antiquark bound state in a color singlet. Mesons made of one or two light quarks are almost all dissociated in $T_c$. Only mesons made of two heavy quarks ($c$, $b$) are bound enough to survive in the range $(1.3-2)$ $T_c$. Although we use a $T$-matrix formulation as well, our parameters have different values of that used in \cite{cabre06} where the main goal was to reproduce mesonic correlators computed on the lattice and not the equation of state. In this last work, the $J/\psi$ meson is bound up to 3 $T_c$ and the $\Upsilon$ meson is bound up to 3.5 $T_c$, thus at much higher $T$ than what we find. It is worth recalling that we are able to compute $T$-matrices in channels where the quark and the antiquark have different masses, which was not considered in \cite{cabre06}.
\item Although strong in the color singlet channel, the contribution of two-body interactions to the equation of state is weak with respect to the free-gas part. This is partly due to a cancellation between attractive and repulsive color channels, that come with an opposite sign in the grand potential. It is tempting to conclude from this result that it provides an \textit{a posteriori} justification of the success of approaches involving free quasiparticles in the description of the equation of state, even in the strongly coupled phase.
\end{itemize}

An obvious drawback of our framework is the neglect of chiral symmetry, leading to results that may be inaccurate in the light quark sector. QCD in Coulomb gauge is currently the formalism which is maximally close to ours while fully including chiral symmetry. Some work has been done in the study of pure Yang-Mills theory and by using a toy model with confining potential that mimics QCD \cite{Lo09}. Modelling the full quark-gluon plasma within Coulomb gauge QCD is however a huge task that still remains to be achieved.

Some comments can be made about the large-$N$ behavior of our results. The meson masses depend on the number of color through the factor $\kappa_{{\bullet; q\bar{q}}}=-\frac{1}{2}(1-\frac{1}{N^2})$ only, so the meson masses are of order 1 at large $N$, with corrections in $1/N^2$ as expected from a quenched potential. Moreover, it has been shown in \cite{LSCB} that the quark contribution to the equation of state behaves as $N_f\, N$ in 't Hooft's limit, as expected. This is an important check of the ability of the present model to deal with the large-$N$ limit.

Future developments of the present model should include the computation of the viscosity-over-entropy ratio. Such a computation can in principle be done without extra parameter. Hence, it is an important extension of our formalism that we hope to present in forthcoming works.

\acknowledgments
G. L. thank F.R.S-FNRS for financial support, and D. Cabrera, R. Rapp and C. Ratti for their interesting discussions and suggestions.
\appendix

\end{document}